\documentclass[twocolumn]{aastex631} 

\usepackage[version=3]{mhchem} 
\usepackage{seqsplit}
\usepackage{enumitem}
\usepackage{tabularx}
\shorttitle{EPACRIS-RMG}
\shortauthors{Yang \& Hu}

\begin{document}

\title{Automated chemical reaction network generation and its application to exoplanet atmospheres}

\correspondingauthor{Jeehyun Yang; Renyu Hu}
\email{jeehyun.yang@jpl.nasa.gov; renyu.hu@jpl.nasa.gov}

\author[0000-0002-1551-2610]{Jeehyun Yang}
\affiliation{Jet Propulsion Laboratory, California Institute of Technology,
Pasadena, CA 91109, USA}

\author[0000-0003-2215-8485]{Renyu Hu}
\affiliation{Jet Propulsion Laboratory, California Institute of Technology,
Pasadena, CA 91109, USA}
\affiliation{Division of Geological and Planetary Sciences, California Institute of Technology, Pasadena, CA 91125, USA}



\begin{abstract}

Along with the advent of JWST and spectroscopic characterization of exoplanet atmospheres with unprecedented detail, there is a demand for a more complete picture of chemical and photochemical reactions and their impact on atmospheric composition. Traditionally, building reaction networks for (exo)planetary atmospheres involves manually tracking relevant species and reactions, a time-consuming and error-prone process. This approach's applicability is also often limited to specific conditions, making it less versatile for different planetary types. (i.e., photochemical networks for Jupiters may not be directly applicable to water-rich exoplanets). We introduce an automated approach using a computer-aided chemical reaction network generator, combined with a one-dimensional photochemical kinetic-transport model, offering significant advantages. This approach automatically selects reaction rates through a rate-based iterative algorithm and multiple refinement steps, enhancing model reliability. Also, this approach allows for the efficient simulation of diverse chemical environments, from hydrogen to water, carbon dioxide, and nitrogen-dominated atmospheres. Using WASP-39b and WASP-80b as examples, we demonstrate our approach's effectiveness, showing good agreement with recent JWST data. Our WASP-39b model aligns with prior studies and JWST observations, capturing photochemically produced sulfur dioxide. The WASP-80b model reveals an atmosphere influenced by deep interior thermochemistry and vertical mixing, consistent with JWST NIRCam observations. Furthermore, our model identifies a novel initial step for the \ce{N2}--\ce{NH3}--\ce{HCN} pathway that enhances the efficiency of the conversion in high-temperature/pressure environments. This automated chemical network generation offers a novel, efficient, and precise framework for studying exoplanetary atmospheres, marking a significant advancement over traditional modeling techniques.

\end{abstract}

\footnote{Supplementary materials: tar.gz file contains (i) RMG input file, (ii) CHEMKIN format file, and (iii) EPACRIS format file.}

\section{Introduction} \label{sec:intro}

Our knowledge of other stellar systems and their accompanying planets has been expanding significantly since the exoplanet surveys of Kepler, K2, and Transiting Exoplanet Survey Satellite (TESS) satellites. Over 5500 exoplanets have been confirmed \citep{NASA_exo_archive}. Adding to this, the recent launch of JWST has provided us with a deluge of high-quality spectroscopic data. This allows for the characterization of exoplanet atmospheres with unprecedented detail, exemplified by the detection of \ce{SO2} in the hot-Jupiter WASP-39~b's atmosphere, which indicates active photochemistry \citep{Tsai_2023}. Another example is the detection of \ce{CO2} in the temperate sub-Neptune K2-18~b's atmosphere \citep{Madhusudhan-2023}, which supports the hypothesis of a water-rich interior \citep{Madhusudhan-2021, Hu_2021b}, but can also be potentially explained by a high-metallicity atmosphere \citep{Yu-2021, Hu_2021b, Tsai_2021, wogan2024jwst}. 

As shown above, JWST enables detailed atmospheric measurements of diverse types of exoplanets from Jupiter-sized to Earth-sized, from cool to hot atmospheres, on which one may expect diverse atmospheric composition and redox conditions. 1-D photochemical atmospheric modeling is crucial for interpreting the JWST observations as well as guiding future observations. Thus, enhancing 1-D photochemical atmospheric modeling with more comprehensive chemical and photochemical reaction networks enables more precise characterization of exoplanet atmospheres, guiding future observations, and advancing our understanding of exoplanets. 

Multiple photochemical reaction networks have already been developed for exoplanet atmospheric modeling studies \citep[e.g.,][]{Moses_2011, Hu_2012, Venot_2012, Tsai_2017, Rimmer2019}. However, there still have been several limitations to constructing these chemical reaction networks applicable to exoplanet atmospheric conditions. One of the major issues is the way photochemical networks are constructed. Most of these existing chemical networks that describe various exoplanet atmospheres are constructed by hand, adopting reaction rate and thermodynamic parameters by carefully keeping track of all possible species and reactions relevant to the target system \citep[e.g.,][]{Moses_2011, Hu_2012, Venot_2012, Tsai_2017, Rimmer2019, Tsai_2021}. This process is very time-consuming and error-prone, and the resulting model significantly depends on the chemistry knowledge of the builder who manually chooses these parameters from various sources (i.e., laboratory measurements, ab initio calculations, estimations, etc.). For this reason, the chance that the number of missing and dubious reactions are included in the model increases as the model size grows, eventually leading to a failure in precisely predicting and interpreting important reaction species and pathways.

A recent study by \cite{Veillet_2023} constructed C-H-O-N chemical networks based on an extensive amount of combustion experimental data gathered over recent decades. This provides a relatively robust chemical network for describing the atmospheres of hot Jupiter exoplanets, predominantly composed of \ce{H2} with an insignificant amount of sulfur species (since the network doesn't contain sulfur-bearing species). However, the applicability of such a model to other types of planets is limited, often necessitating significant time and effort to develop another model for another system. This limitation arises because the relevance of specific chemical species and reactions is intrinsically tied to the system's conditions, such as temperature, pressure, and the dominant gas species. As a result, the chemical network built by \cite{Veillet_2023}, while built with an intent to model \ce{H2}-dominated atmospheres only, is not applicable for, e.g., Venus-like exoplanet atmospheres, given the differences in temperature and pressure profiles, as well as in the dominant atmospheric gas composition (e.g., \ce{CO2}-dominated atmosphere with sulfur-bearing species). While some might suggest including all known species and reactions studied so far, doing so is impractical. The computational time required for large chemical kinetics simulations scales approximately linearly with the number of chemical reactions and approximately quadratically with the number of chemical species, \textit{N} \citep{Schwer-2002}. Given the vast amount of spectroscopic data expected from the JWST and future observational missions, a fundamentally new approach to photochemical reaction network construction is essential.

Over the past decade, advancements in computational chemical engineering have paved the way for automated chemical reaction network generation. These automation techniques can be categorized based on their approaches to species and reaction selection, as well as parameter generation. One common method is to define reaction families to find possible reactions, which allows for the expansion of the network starting from an initial set of molecules \citep[e.g.,][]{Sarathy_2012}. Another approach is a rule-based method, as adopted by packages like Genesys \citep[e.g.,][]{Vandewiele_2012}. The inclusion of species and reactions in the network is determined by a set of user-defined constraints. Although computationally efficient, these constraints, often dependent on the user's chemistry knowledge, can potentially bias network generation. In contrast, the Reaction Mechanism Generator (RMG, \cite{Gao_2016,rmg-v3, RMG-database}) employs a rate-based method, where the importance of a species or reaction is determined based on iterative simulations of the chemical system. For this reason, this rate-based approach is more objective (i.e., independent of the user's chemistry knowledge) and can provide a more comprehensive chemical network than other methods. The only downside of this rate-based method is that it is computationally more expensive than other methods. These RMG-generated networks have been actively utilized in various chemical engineering fields, such as a computer-generated acetylene pyrolysis model by \cite{Liu_2020}, which successfully described the previous experiment by \cite{Norinaga_2008}, butyl acetate pyrolysis and combustion model by \cite{Dong-2023}, and methyl propyl ether pyrolysis and oxidation model by \cite{johnson-2021} showing excellent agreement with most of the shock tube and rapid compression machine data. Notably, this automation has recently been used to simulate and successfully rationalize the previous laboratory photochemical studies by \cite{Fleury_2019, Fleury_2020} that simulated hot Jupiter-like atmospheric conditions \citep{Yang-2023}. Such applications underscore the reliability and vast potential of automated reaction network generators, particularly when used together with existing photochemical reaction networks, offering solutions to challenges associated with manual methodologies.

Given the challenges and potential of recent advancements in computational chemical engineering, we develop the atmospheric chemistry module of EPACRIS (ExoPlanet Atmospheric Chemistry \& Radiative Interaction Simulator), an innovative atmospheric simulation framework for exoplanets. The atmospheric radiative transfer module of EPARCRIS will be described in a separate paper (Scheucher \& Hu, in prep). The EPACRIS atmospheric chemistry module integrates a cutting-edge automated chemical reaction network generation by RMG with a general-purpose one-dimensional photochemical kinetic-transport atmospheric simulation, originally developed by \cite{Hu_2012}, and since then expanded and upgraded substantially \citep{Hu_2013, Hu_2014, Hu_2019, Hu_2021}. This integration facilitates the fast and reliable construction of tailored reaction networks for specific exoplanet atmospheres. This paper details our methodology and demonstrates its effectiveness using the well-characterized atmosphere of WASP-39~b and the atmosphere of WASP-80~b as case studies for two different types of \ce{H2}-dominated hot or warm Jupiters, compared with the recent JWST observations and photochemical modeling studies \citep{Alderson-2023, Ahrer-2023, Feinstein-2023, jtec2023, Rustamkulov-2023, Tsai_2023, Powell_2024, Bell-2023}.

\begin{figure*}
    \centering
    \includegraphics[width=1\textwidth]{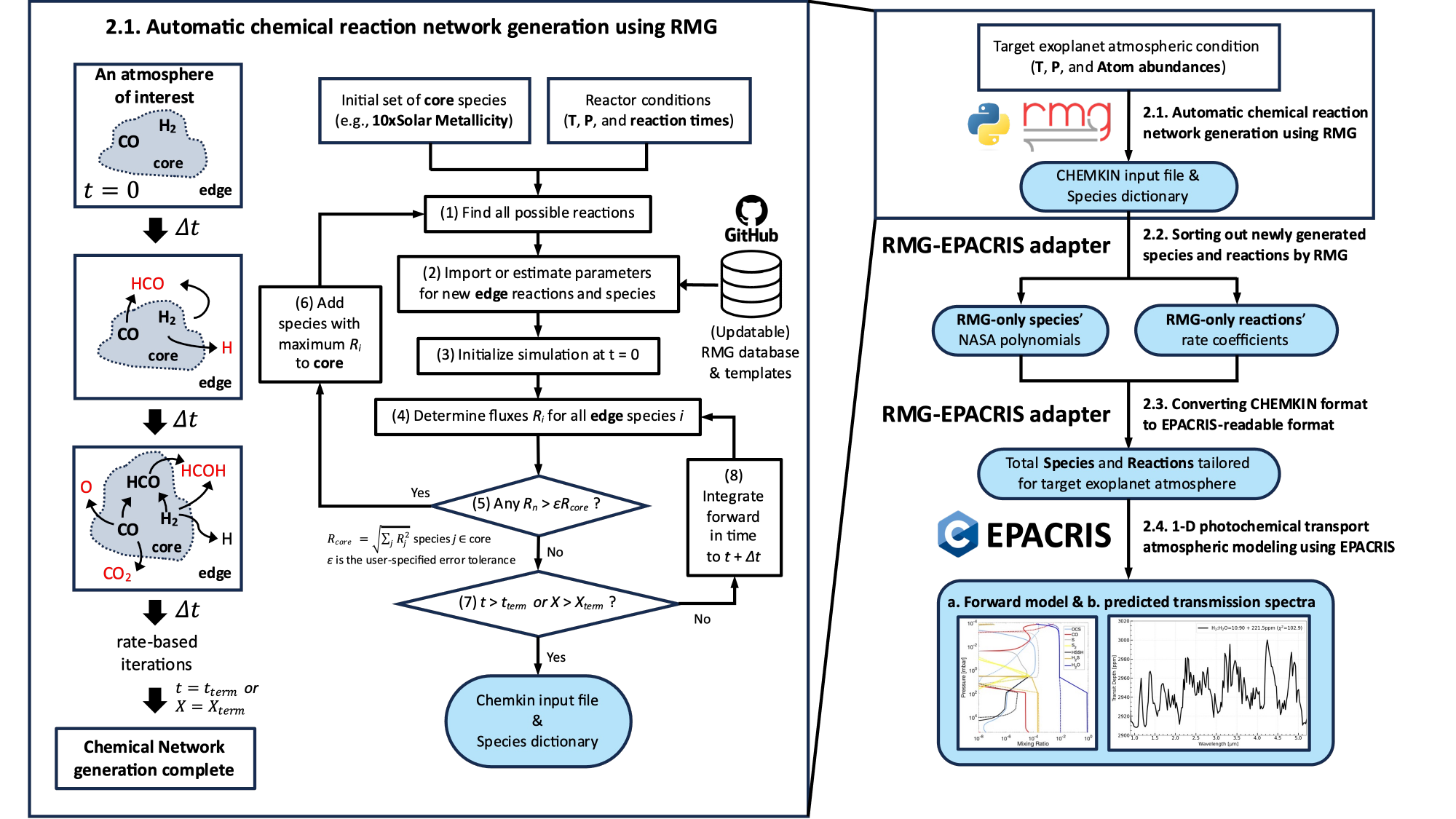}
    \tablecomments{\footnotesize\ Mention of any commercial product, process, or service by name, trademark, or manufacturer does not imply endorsement by the U.S. Government or the Jet Propulsion Laboratory, California Institute of Technology}
    \caption{\footnotesize A schematic diagram and the flowchart describing the expansion of the chemical network during automated reaction network generation by RMG using the rate-based algorithm (left) and overall workflow of implementation into 1-D photochemical transport atmospheric modeling (right) in this work. RMG stands for the Python-written ``Reaction Mechanism Generator'' \citep{Gao_2016, rmg-v3, RMG-database}, while EPACRIS stands for the overall ``ExoPlanet Atmospheric Chemistry \& Radiative Interaction Simulator'' written in C. Each light blue-colored shaded box refers to the corresponding output after each method described in Section \ref{sec:methods}.}
    \label{fig:method-workflow}
\end{figure*}

\section{Methods} \label{sec:methods}
A schematic diagram of the methodology and overall workflow adopted in this study is reported in Figure \ref{fig:method-workflow}.

\subsection{Automatic chemical reaction network generation using the Reaction Mechanism Generator (RMG) software}\label{sec:rmg}

A detailed chemical reaction network for modeling the \ce{H2}-dominated atmospheres of warm and hot Jupiters with equilibrium temperatures of 800--1500 K ($T_{\rm eq}$ of WASP-39~b and WASP-80~b are within this range) was constructed automatically by RMG \citep{Gao_2016, RMG-database}. RMG is a Python-based open-source software and has been extensively used in the chemical engineering community to automatically generate chemical networks to simulate numerous pyrolysis and combustion chemistry successfully \citep{Class-2016, Dana-2019, Chu-2019, Keceli-2019, Liu_2020}. RMG is previously described in detail in \cite{Gao_2016, rmg-v3, RMG-database} and only briefly described here along with Figure~\ref{fig:method-workflow}.

In given reactor conditions (i.e., temperature, pressure, reaction time, and initial mixing ratio of gas species), RMG will first place the initial species in the reaction system into the `core' of the model and then find all the possible reactions based on these `core' species (i.e., indicated as (1) on the left side of Figure~\ref{fig:method-workflow}). Chemical reaction rates depend on the species concentrations at the previous timestep. Thus, determining this initial set of core species is crucial when applying automatic reaction mechanism generation to (exo)planetary atmospheres. Unlike laboratory experiments, where initial concentrations are well-controlled, such information is often not fully available for exoplanet atmospheres. In response to this challenge, there are several ways to address this issue.

One approach to setting initial conditions is based on existing observational constraints. For instance, the recent JWST observation of K2-18~b has constrained concentrations of certain chemical species, such as methane (\ce{CH4}) and carbon dioxide (\ce{CO2}), from atmospheric retrievals \citep{Madhusudhan-2023}. Building on this, researchers can construct the chemical network of interest by assuming specific compositions—e.g., approximately 50\% \ce{H2O} and 50\% \ce{H2}, along with constraints on \ce{CH4}, \ce{CO2}, and \ce{(CH3)2S}), which is tailored to explore the "Hycean Worlds" scenario of K2-18~b proposed by \cite{Madhusudhan-2021}.

Alternatively, one can define a set of grids, such as varying solar metallicity \citep{Lodders-2020} from 1$\times$to 100$\times$ solar metallicity as inputs for Reaction Mechanism Generator (RMG) and build a chemical network for different solar metallicities. Benchmarking these networks against existing JWST observations can check the feasibility of each chemical network.  In the context of planetary atmospheres, constructing a chemical network tailored for a \ce{CO2}-dominated atmosphere with a trace amount of sulfur species can provide insights into the atmospheres of Venus or Venus-like exoplanets. Such applications offer invaluable insights into (exo)planetary atmospheres.

After the reactor condition is provided, the next step for RMG (indicated as (2) in Figure~\ref{fig:method-workflow}) is to simulate the possible reactions using its database (maintained and updated with the latest data sources by the developers of \cite{RMG-developers}) of reaction parameters from previous experiments, ab initio calculations, or estimation methods (e.g., Benson group additivity, \cite{Benson_2004}), which will generate a list of possible product species (i.e., `edge' species). It should be noted that RMG relies on a chemical kinetics database compiled from various sources, each with its inherent errors. However, a well-maintained database represents our best knowledge at any given time. RMG initializes simulation at t=0 (indicated as (3) in Figure \ref{fig:method-workflow}), followed by the next steps (indicated as (4--6) in Figure \ref{fig:method-workflow}) that determine if these `edge' species are important enough to be added to the `core' species. `Edge' species \textit{i} are included into the `core' species if
\begin{equation}
    R_{i} = \frac{dC_i}{dt}>\epsilon R_{core} \label{eq:edge_to_core},
\end{equation} 
where $R_{i}$ is the production and loss flux of `edge' species \textit{i}, defined as an infinitesimal change in the concentration of `edge' species (i.e., $dC_{i}$) in an infinitesimal time (i.e., $dt$), $\epsilon$ is the user-specified error tolerance, and $R_{core}$ is the characteristic flux of the reaction system, defined by
\begin{equation}
    R_{core} = \sqrt{\sum_{j}R_j\textsuperscript{2}} \quad \text{species \textit{j} $\in$ core.}
    \label{eq: characteristic flux}
\end{equation} 
A typically recommended range for this user-specified error tolerance, $\epsilon$, is between 0.01 and 0.05 for users seeking a larger and more comprehensive model, despite the higher computational cost. Consequently, $\epsilon$ was set to 0.1 in this work (as specified by \textbf{\texttt{toleranceMoveToCore}=0.1} in the RMG input file, available in the supplementary materials).
As shown in (7--8) in Figure \ref{fig:method-workflow}, the reaction generation and integration steps continue until they meet the termination criteria for reaction time, $t_{term}$ or the concentration of a specific species $X_{term}$. This process results in the completed chemical network, encompassing all `core' species and reactions with significant fluxes at the given reactor conditions. 

In this work, temperatures from 700 to 2000 K and pressure between 10\textsuperscript{-3} and 10\textsuperscript{2} bar were sampled to generate chemical networks relevant within these \textit{T} and \textit{P} ranges using the ranged reactors setting in RMG \citep{rmg-v3}, later then combined. We used an initial molecular mixing ratio of 10$\times$solar metallicity, following the previous models of WASP-39~b \citep{Tsai_2023} and WASP-80~b \citep{Bell-2023}, and automatically generated a chemical network. The reaction time criterion, \textit{$t_{term}$}, was set to 3.154$\times$10\textsuperscript{16} s, or 10\textsuperscript{9} years, assuming the reaction time required for the reaction to achieve chemical equilibrium. The choice of reaction libraries (\texttt{\seqsplit{Klippenstein\_Glaborg2016, primarySulfurLibrary, primaryNitrogenLibrary, NOx2018,}} and \texttt{\seqsplit{Nitrogen\_Glaborg\_Gimenez\_et\_al}}) and thermochemical libraries (\texttt{\seqsplit{SABIC\_aromatics, primaryThermoLibrary, BurkeH2O2, thermo\_DFT\_CCSDF112\_BAC, DFT\_QCI\_thermo, Klippenstein\_Glaborg2016, CH, primaryNS}} and \texttt{\seqsplit{SulfurGlarborgMarshall}}) from which RMG retrieves rate parameters during chemical network generation can be found in the RMG input file (see supplementary materials), and the details of these libraries (e.g., rate constants, references, etc.) can be found in the RMG database \citep{RMG-developers}. The pressure dependence feature of RMG was enabled to automatically construct pressure-dependent networks for species with up to 10 atoms (i.e., constraining a total number of atoms). The resulting chemical network contained 105 species and 2337 reactions (forward-reverse reaction pairs), which can be found in the supplementary information as the CHEMKIN input file. Among these 2337 generated reactions, 2271 reactions didn't violate their respective collision limits, $k_{\text{coll}}$ (i.e., any bimolecular reaction rate coefficient doesn't exceed its Lennard-Jones collision rate constant), which were then incorporated into our 1D kinetic-transport model (Section~\ref{sec:epacris}) after the adaptation described in Sections~\ref{sec:sorting} and \ref{sec:RMG-EPACRIS_conversion}.

\begin{figure}
    \centering
    \includegraphics[width=0.5\textwidth]{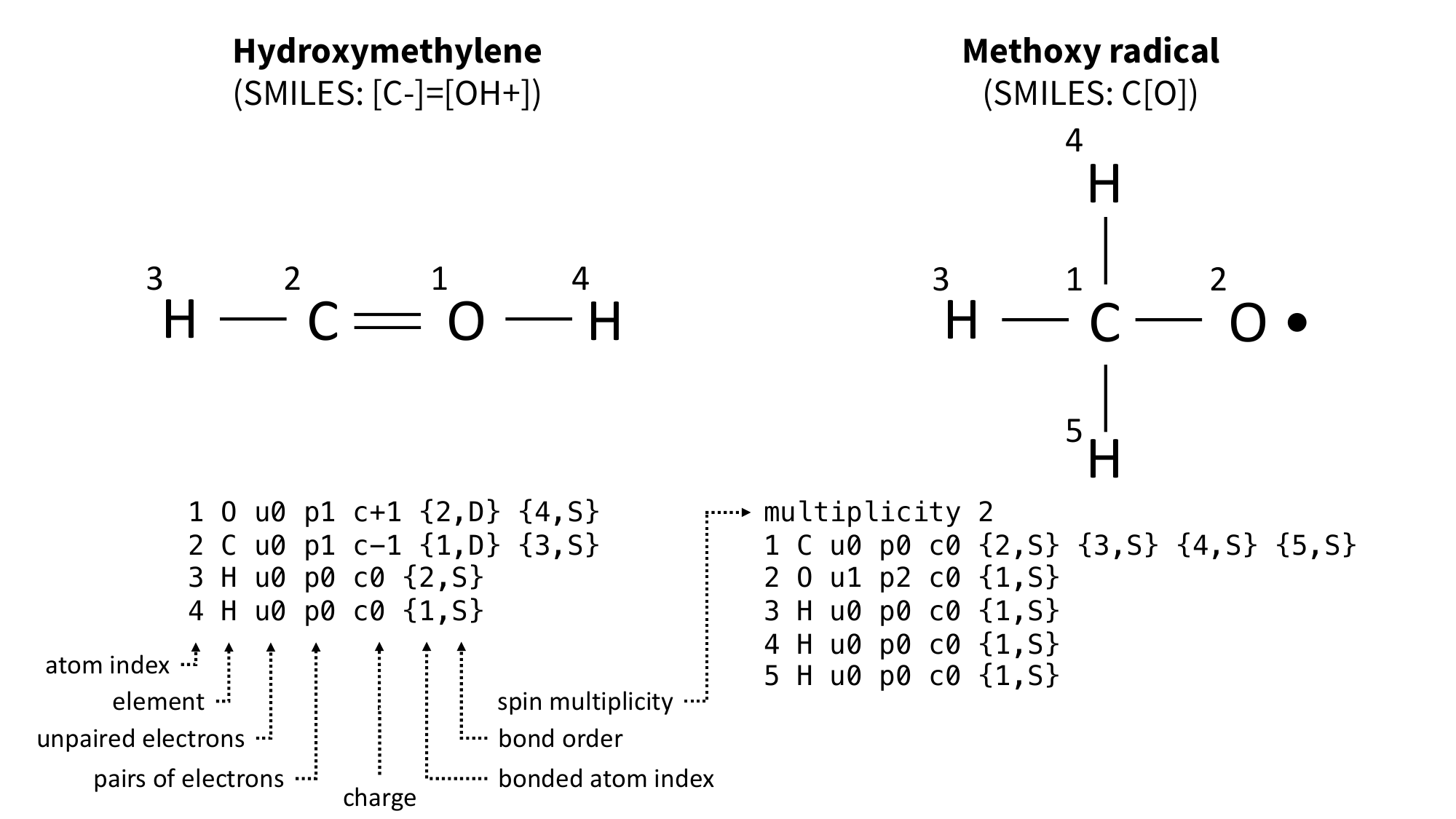}
    \caption{\footnotesize Adjacency list representations of hydroxymethylene (left) and methoxy radical (right). SMILES stands for Simplified Molecular-Input Line-Entry System.}
    \label{fig:adjacency_list}
\end{figure}

\begin{table*}[hb]
\caption{111 molecular species originally included in the EPACRIS library}
\centering
\begin{tabular}{|ll|ll|ll|}
\hline
\textbf{Species} & \textbf{SMILES}\textsuperscript{\textit{a}} & \textbf{Species} & \textbf{SMILES}\textsuperscript{\textit{a}} & \textbf{Species} & \textbf{SMILES}\textsuperscript{\textit{a}}\\ \hline

   \ce{O}\textsuperscript{\textit{b}} & [O]&\ce{O3}\textsuperscript{\textit{d}}& [O-][O+]=O & \ce{H}\textsuperscript{\textit{b}} & [H]\\ \hline
   \ce{OH}\textsuperscript{\textit{b}}& [OH] & \ce{HO2}\textsuperscript{\textit{b}} & [O]O&\ce{H2O2}\textsuperscript{\textit{b}}& OO \\ \hline   
   \ce{H2O}\textsuperscript{\textit{b}} & O &\ce{N}\textsuperscript{\textit{b}}& [N] & \ce{NH3}\textsuperscript{\textit{b}} & N \\ \hline
   \ce{NH2}\textsuperscript{\textit{b}}& [NH2]&\ce{N2O}\textsuperscript{\textit{b}}&[N-]=[N+]=O&\ce{NO}\textsuperscript{\textit{b}}& [N]=O \\ \hline   
   \ce{NO2}\textsuperscript{\textit{b}} & [O-][N+]=O&\ce{NO3}\textsuperscript{\textit{d}}& [O-][N+](=O)[O] &
   \ce{N2O5}\textsuperscript{\textit{d}} & [O-][N+](=O)O[N+](=O)[O-]\\ \hline   
   \ce{HNO}\textsuperscript{\textit{b}}& N=O & \ce{HNO2}\textsuperscript{\textit{b}} & [O-][NH+]=O&\ce{HNO3}\textsuperscript{\textit{d}}& [O-][N+](=O)O \\ \hline   
   \ce{C}\textsuperscript{\textit{b}} & [C]&\ce{CO}\textsuperscript{\textit{b}}& [C-]\#[O+] & \ce{CH4}\textsuperscript{\textit{b}} & C \\ \hline 
   \ce{CH2O}\textsuperscript{\textit{b}}& C=O &\ce{CH2O2}\textsuperscript{\textit{b}}&O=CO& \ce{CH3OH}\textsuperscript{\textit{b}} & CO  \\ \hline   
  \ce{CH3OOH}\textsuperscript{\textit{d}}& COO &\ce{C2}\textsuperscript{\textit{d}} & [C]\#[C]& \ce{C2H2}\textsuperscript{\textit{b}}& C\#C \\ \hline
  \ce{C2H3}\textsuperscript{\textit{b}} & [CH]=C&\ce{C2H4}\textsuperscript{\textit{b}}& C=C &\ce{C2H5}\textsuperscript{\textit{b}}&C[CH2] \\ \hline  
  \ce{C2H6}\textsuperscript{\textit{b}}& CC & \ce{HCCO}\textsuperscript{\textit{b}} & [CH]=C=O&\ce{CH2CO}\textsuperscript{\textit{b}}&C=C=O\\ \hline
  \ce{CH3CO}\textsuperscript{\textit{b}}&C[C]=O& \ce{CH3CHO}\textsuperscript{\textit{b}} & CC=O&\ce{C2H5O}\textsuperscript{\textit{d}}& CC[O]\\ \hline  
  \ce{HCN}\textsuperscript{\textit{b}} & C\#N &\ce{CN}\textsuperscript{\textit{b}}& [C]\#N &\ce{NCO}\textsuperscript{\textit{b}} & [O]C\#N\\ \hline
  \ce{S}\textsuperscript{\textit{b}}& [S] &\ce{S2}\textsuperscript{\textit{b}} & [S][S]&\ce{SO}\textsuperscript{\textit{b}}& [S][O] \\ \hline
   \ce{SO2}\textsuperscript{\textit{b}} & O=S=O&\ce{SO3}\textsuperscript{\textit{b}}& O=S(=O)=O & \ce{H2S}\textsuperscript{\textit{b}} & S \\ \hline
   \ce{SH}\textsuperscript{\textit{b}}& [SH] & \ce{HSO}\textsuperscript{\textit{b}} & O=[SH]&\ce{HOSO}\textsuperscript{\textit{b}}& O[S]=O \\ \hline
   \ce{OCS}\textsuperscript{\textit{b}} & O=C=S&\ce{CS}\textsuperscript{\textit{d}}& [C-]\#[S+] & \ce{CH3S}\textsuperscript{\textit{b}} & C[S] \\ \hline
   \ce{CO2}\textsuperscript{\textit{b}}& O=C=O & \ce{H2}\textsuperscript{\textit{b}} & [H][H]&\ce{O2}\textsuperscript{\textit{b}}& [O][O] \\ \hline
   \ce{N2}\textsuperscript{\textit{b}} & N\#N&\ce{O}(\textsuperscript{1}D)\textsuperscript{\textit{d}}& O\textsuperscript{\textit{c}} 
   &\ce{NH}\textsuperscript{\textit{b}} & [NH]\\ \hline
   \ce{CH}\textsuperscript{\textit{b}}& [CH] & \ce{CH2}\textsuperscript{\textit{b}} & [CH2]&\ce{CH3}\textsuperscript{\textit{b}}& [CH3] \\ \hline
   \ce{HCO}\textsuperscript{\textit{b}} & [CH]=O&\ce{CH3O}\textsuperscript{\textit{b}}& C[O]& \ce{HOCO}\textsuperscript{\textit{b}} & O=[C]O  \\ \hline
   \ce{C2H}\textsuperscript{\textit{b}}& [C]\#C & \ce{CH3NO2}\textsuperscript{\textit{d}} & [O-][N+](=O)C&\ce{CH3NO3}\textsuperscript{\textit{d}}& [O-][N+](=O)OC \\ \hline
   \ce{CH2CN}\textsuperscript{\textit{b}} & C=C=[N]&\ce{HOSO2}\textsuperscript{\textit{b}}& [O]S(=O)O & \ce{CH3SH}\textsuperscript{\textit{b}} & CS \\ \hline
   \ce{HNO4}\textsuperscript{\textit{d}}& [O-][N+](=O)OO & \ce{CH3OO}\textsuperscript{\textit{b}} & CO[O]&\ce{HNCO}\textsuperscript{\textit{b}}& N=C=O \\ \hline
   \ce{H2SO4}\textsuperscript{\textit{d}} & OS(=O)(=O)O&\textsuperscript{1}\ce{SO2}\textsuperscript{\textit{d}}& [O]S[O]\textsuperscript{\textit{c}} &
   \textsuperscript{3}\ce{SO2}\textsuperscript{\textit{d}} & [O]S[O]\\ \hline
   
   \ce{S3}\textsuperscript{\textit{d}} & S=S=S&\ce{S4}\textsuperscript{\textit{d}}& S=S=S=S &\ce{H2SO4}(A)\textsuperscript{\textit{d}} & aerosol\\ \hline
   \ce{S8}\textsuperscript{\textit{d}} & S1SSSSSSS1&\textsuperscript{1}\ce{CH2}\textsuperscript{\textit{b}}& [CH2]\textsuperscript{\textit{c}} &\ce{C3H2} & 	[CH]C\#C\\ \hline
   \ce{H2CCCH} & [CH]=C=C&\ce{H3CCCH}& C\#CC &\ce{H2CCCH2} & C=C=C\\ \hline
   \ce{C3H5} & [CH2]C=C&\ce{C3H6}& C=CC &\ce{C3H7} & [CH2]CC\\ \hline
   \ce{C3H8} & CCC&\ce{C4H}& [C]\#CC\#C &\ce{C4H2} & C\#CC\#C\\ \hline
   \ce{C4H3} & [CH]=CC\#C&\ce{C4H4}& C\#CC=C &\ce{C4H5} & C=[C]C=C\\ \hline
   \ce{CH3CH2CCH} & C\#CCC&\ce{CH3CHCCH2}& C=C=CC &\ce{CH2CHCHCH2} & C=CC=C\\ \hline
   \ce{C4H8} & C=CCC&\ce{C4H9}& [CH2]CCC &\ce{C4H10} & CCCC\\ \hline
   \ce{C6H} & [C]\#CC\#CC\#C&\ce{C6H2}& C\#CC\#CC\#C & \ce{C6H3} & C\#CC\#C[C]=C\\ \hline
   \ce{C6H6} & c1ccccc1&\ce{C8H2}& C\#CC\#CC\#CC\#C & \ce{CMHA}& OCC1OC(O)C(O)C(O)C1O\\ \hline
   \ce{N2H2}\textsuperscript{\textit{b}} & N=N&\ce{N2H3}\textsuperscript{\textit{b}}& [NH]N & \ce{N2H4}\textsuperscript{\textit{b}} & NN\\ \hline
   \ce{CH3NH2}\textsuperscript{\textit{b}} & CN &\ce{CH3CHNH}& CC=N & \ce{S8}(A)\textsuperscript{\textit{d}} & aerosol\\ \hline

\end{tabular}
\label{tbl: originalspecies}
\tablecomments{\footnotesize\textsuperscript{\textit{a}} Simplified Molecular-Input Line-Entry System\\ 
\textsuperscript{\textit{b}} 65 chemical species that were included in the original EPACRIS library and were also picked by RMG to be important for describing atmospheric conditions of WASP-39~b and WASP-80~b.\\
\textsuperscript{\textit{c}} Despite appearing as other chemical species or different spin states in SMILES representation, these species are singlets in the `adjacency lists' representation, indicating all electrons are paired.\\
\textsuperscript{\textit{d}} 20 chemical species additionally included in the photochemical network to fully account for photochemistry and aerosol chemistry that might be important in the atmospheres of WASP-39~b and WASP-80~b.}
\end{table*}

\begin{table}[hb]
\caption{40 newly included molecular species in the chemical network tailored for \ce{H2}-dominated atmospheres by RMG}
\centering
\begin{tabular}{|ll|ll|}
\hline
\textbf{Species} & \textbf{SMILES}\textsuperscript{\textit{a}} & \textbf{Species} & \textbf{SMILES}\textsuperscript{\textit{a}}\\ \hline
   \ce{CH2OH} & [CH2]O&\ce{HCOH}& [C-]=[OH+] \\ \hline
   \ce{CH2CHO} &[CH2]C=O&\ce{H2CC}&[C]=C\\ \hline
   \ce{CHCHO}&[CH]=C[O]&c\ce{C2H3O}&[CH]1CO1\\ \hline
   \ce{OCHCO} &O=[C]C=O&\ce{HOCH2O}&[O]CO\\ \hline
   \ce{OCHO} &[O]C=O&\ce{HSO2}&O=[SH]=O\\ \hline
   \ce{HOS} &O[S]&\ce{S}a&[S]\textsuperscript{\textit{b}}\\ \hline
   \ce{HSS}&S=[SH]&\ce{HSSH}&SS\\ \hline
   \ce{CH2SH}&[CH2]S&\ce{HCCS}&[S]C\#C\\ \hline
   \ce{H2SS}&S=[SH2]&\ce{H2CN}&C=[N]\\ \hline
   \ce{C2N2}&N\#CC\#N&\ce{CH3CN}&CC\#N\\ \hline
   \ce{CH2NH} &C=N&\ce{NCOH}&OC\#N\\ \hline
   \ce{NNH}&[N]=N&\ce{NH2NO}&NN=O\\ \hline
   \ce{HNOH}&[NH]O&\ce{HONO}&ON=O\\ \hline
   \ce{H2NO}&N[O]&\ce{NH2OH}&NO\\ \hline
   \ce{H2NONO}&NON=O&\ce{HNO}(T)&[NH][O]\\ \hline
   \ce{CH3NH}&C[NH]&\ce{CH2NH2}&[CH2]N\\ \hline
   \ce{CHNH}&[CH]=N&\ce{HNC}-2&[C-]\#[NH+]\\ \hline
   \ce{NCCN}&[N]=C=C=[N]&\ce{H2NCHO}&NC=O\\ \hline
   \ce{H2NCO}&N[C]=O&\ce{CH3CHN}&CC=[N]\\ \hline
   \ce{C2H5CO}&CC[C]=O&\ce{NH}-2&[NH]\textsuperscript{\textit{b}}\\ \hline
\end{tabular}
\label{tbl: newspecies}
\tablecomments{\footnotesize\textsuperscript{\textit{a}} Simplified Molecular-Input Line-Entry System\\ \textsuperscript{\textit{b}} Despite appearing as doublet or triplet radicals in SMILES representation, these species are singlets in the `adjacency lists' representation, indicating all electrons are paired.}
\end{table}

\subsection{Sorting out newly generated species and reactions by RMG}\label{sec:sorting}

Besides the RMG-generated reactions and species outlined in Section \ref{sec:rmg}, our initial kinetic-transport atmosphere model already possesses a reaction library that includes photodissociation and associated reactions. This library comprises 111 species (see Table~\ref{tbl: originalspecies}) and 914 reactions, broken down as follows: 657 bi-molecular reactions, 91 ter-molecular reactions, 93 thermo-dissociation reactions, and 71 photochemistry reactions \citep{Hu_2012,Hu_2014,Hu_2021}. The rates of photochemical reactions are calculated according to Equation (12), as detailed in Section 2.3 of \cite{Hu_2012}. Because RMG does not generate photochemistry-driven reactions, it is necessary to combine the original reaction network and the network generated by RMG, and the first step is to identify any overlapping species and reactions to prevent duplicates. We have annotated the 111 original species using RMG's `adjacency lists' methodology \citep{Gao_2016, RMG-database}, which allows the RMG-EPACRIS adapter to compare the reactants and products, including reverse reactions, and ensure no duplications. The `adjacency lists' method uses a graph-based structure to illustrate molecules, identifying atoms as vertices and their connecting bonds as edges in the list. For example, the adjacency list for hydroxymethylene (HCOH) and methoxy radical (\ce{CH3O}) are shown in Figure \ref{fig:adjacency_list}. The structure of the `adjacency lists' method is defined as follows: The first column specifies the atom index, the second column specifies the atom element, and the third column, prefixed by the lowercase letter `u' for ``unpaired'', specifies the count of unpaired electrons for each atom. The fourth column, prefixed by the lowercase letter `p' for ``pairs'', specifies the number of lone electron pairs. The fifth column, prefixed by the lowercase letter `c' for ``charge,'' specifies the formal charge on the atom. Bracketed values specify the presence of a bond, with the first value (i.e., number) indicating the index of the atom to which the current atom is bonded, and the second value (i.e., the uppercase letter) denoting the bond order: `S' for single, `D' for double, `T' for triple, or `B' for benzene-type bonds. If the molecule has an overall spin multiplicity (i.e., the degeneracy of the electronic ground state) larger than 1, it will be defined above the adjacency list (e.g., see the methoxy radical case in Figure \ref{fig:adjacency_list}). In the adjacency list of the methoxy radical molecule shown on the right side of Figure~\ref{fig:adjacency_list}, the oxygen atom has a single unpaired electron (thus having an overall spin multiplicity of 2) and 1 single bond to the carbon atom that has 3 single bonds to hydrogen atoms, forming a methoxy radical.
After sorting out newly generated species and reactions by RMG using the adjacency lists method, we found that 65 species (indicated with the footnote \textit{b} in Table~\ref{tbl: originalspecies}) were both included in the RMG-generated network and the original EPACRIS library, with 40 new species generated by RMG as shown in Table~\ref{tbl: newspecies}. We included non-reactive species, helium and Neon, and 20 additional chemical species (indicated with the footnote \textit{d} in Table~\ref{tbl: originalspecies}) as well as their other relevant thermochemical reactions imported from the original EPACRIS reaction library in the photochemical network. We found that the species that were not included in the RMG-generated list (i.e., thermochemically not important) are mostly associated with photodissociation (e.g., O(\textsuperscript{1}D)) and the chemical network enabled by photodissociation. Consequently, we included them in our analysis to account for the impacts of photochemistry and aerosol chemistry, which might be significant in the atmospheres of WASP-39~b and WASP-80~b, except for the molecules that have more carbon atoms than \ce{C3} hydrocarbons. Although \ce{C3} and larger species are observed in Titan's atmosphere and mainly formed through photochemistry \citep{Yung-1984}, we omitted them here due to the unfavorable physical (high temperature) and chemical conditions (\ce{H2}-dominated) in hot Jupiters, and the uncertainties in relevant photodissociation rates. 
As a result, the final photochemical network contained 126 species and 2578 reactions (693 original reactions = 71 photochemistry reactions + 529 bi-molecular reactions + 58 ter-molecular reactions + 35 thermo-dissociation reactions, and 1885 reactions newly generated by RMG). Except for the 71 photochemistry reactions, the other 2484 reactions are forward-reverse reaction pairs. In the future, we will incorporate only the thermal-driven reaction rate coefficients that are generated by RMG, and only the photochemical reaction rate coefficients that are stored in the original reaction list into 1D kinetic-transport modeling.

\subsection{Converting CHEMKIN format to EPACRIS-readable format}\label{sec:RMG-EPACRIS_conversion}

After automatic chemical network generation by RMG as described in Section \ref{sec:rmg}, the thermodynamic parameters (i.e., heat capacity, enthalpy, and entropy) and rate coefficients are provided in both CHEMKIN and Cantera \citep{cantera} formats. The RMG-EPACRIS adapter then imports this information and converts it into a format adopted by EPACRIS to become inputs to 1D photochemical kinetic-transport atmospheric models, as described in the following Sections~\ref{sec:thermo} and \ref{sec:kinetics}.

\subsubsection{Thermodynamic parameters}\label{sec:thermo}

RMG uses the NASA polynomial representation \citep{NASA_poly} to calculate the relevant thermodynamic parameters. The NASA polynomial representation was originally developed by scientists at NASA to express temperature-dependent thermodynamic parameters such as the heat capacity \textit{C$_p$}(\textit{T}), enthalpy \textit{H}(\textit{T}), and entropy \textit{S}(\textit{T}) using seven or nine coefficients \citep{NASA_poly}. In this representation, the following thermodynamic parameters are given by
nine polynomial coefficients \textbf{\textit{a}} = [\textit{a$_{-2}$, a$_{-1}$, a$_{0}$, a$_{1}$, a$_{2}$, a$_{3}$, a$_{4}$, a$_{5}$, a$_{6}$}] ($a_{-2}=a_{-1}=0$ \text{in the seven-coefficient version}):
\begin{equation}
    \begin{split}
    C_p(T) = & R(a_{-2}T^{-2}+a_{-1}T^{-1}+a_{0}+a_{1}T+a_{2}T^{2}\\    &+a_{3}T^{3}+a_{4}T^{4})\label{eq:heat_capacity}
    \end{split}
\end{equation}

\begin{equation}
    \begin{split}
    H(T) = &
    R(-a_{-2}T^{-1}+a_{-1}\text{ln} T+a_{0}T+\frac{1}{2}a_{1}T^2+\\
    &\frac{1}{3}a_{2}T^3+\frac{1}{4}a_{3}T^4+\frac{1}{5}a_{4}T^5+a_{5})\label{eq:enthalpy}
    \end{split}
\end{equation}

\begin{equation}
    \begin{split}
    S(T) = &R(-\frac{1}{2}a_{-2}T^{-2}-a_{-1} T^{-1}+a_{0}\text{ln}T+a_{1}T+\\
    &\frac{1}{2}a_{2}T^2+\frac{1}{3}a_{3}T^3+\frac{1}{4}a_{4}T^4+a_{6})
    \label{eq:entropy}
    \end{split}
\end{equation}
Then, the RMG-EPACRIS adapter obtains each species' Gibbs free energy, \textit{G}, by the following equation:
\begin{equation}
    G(T) = H(T)-TS(T)
    \label{eq:gibbs_free_energy}
\end{equation}
The Gibbs free energy of species is used to calculate the reverse reaction rates in 1D photochemical kinetic-transport models using the methods outlined in \cite{Hu_2014}.

\subsubsection{Rate-coefficient expressions}\label{sec:kinetics}

RMG adopts eight types of expressions for reaction rate constants. The RMG-EPACRIS adapter converts these into the formats adopted by EPACRIS (available in supplementary materials), enabling the importation of the rate constants, $k$, for 1D kinetic-transport atmospheric modeling. Consequently, EPACRIS implements the same eight rate-coefficient expressions, as elaborated below.

\begin{itemize}[leftmargin=*, label={}]

\item[]\textbf{[1] Arrhenius}-type expression

\vspace{0.5mm} 

Type 1 is the Arrhenius-type expression whose temperature-dependent rate-coefficient, $k(T)$, follows the Arrhenius equation as shown in Equation \ref{eq:arrhenius}.
\begin{equation}
    k(T) = A\left({\frac{T}{T_0}}\right)^n\text{exp}\left(-\frac{E_a}{RT}\right).
    \label{eq:arrhenius}
\end{equation}
In the Arrhenius equation, $A$ represents the pre-exponential factor, $T_0$ the reference temperature in kelvins [K], $n$ the temperature exponent, and $E_0$ the activation energy in joules per mole [$J\cdot mol^{-1}$]. Here, $T$ denotes temperature [K], and $R$ is the ideal gas constant, 8.314 [$J \cdot mol^{-1} \cdot K^{-1}$]. The unit of $A$ depends on the reaction order—[$s^{-1}$] for first-order (i.e., unimolecular reaction, an elementary reaction in which the rearrangement of a single reactant produces one or more products), [$m^3 \cdot mol^{-1} \cdot s^{-1}$] for second-order (i.e., bimolecular reaction, involving the simultaneous collision of any combination of two reactants), and [$m^6 \cdot mol^{-2} \cdot s^{-1}$] for third-order (i.e., termolecular reaction, an elementary reaction involving the simultaneous collision of any combination of three reactants) reactions.

\item[]\textbf{[2] Multi-Arrhenius}-type expression

\vspace{0.5mm} 

Type 2 is the Multi-Arrhenius-type expression whose temperature-dependent rate-coefficient, $k(T)$, follows a set of Arrhenius equations summed to obtain the overall rate-coefficient, as shown in Equation \ref{eq:multiarrhenius}.
\begin{equation}
    k(T) = \sum_{i}A_i\left({\frac{T}{T_{0,i}}}\right)^{n_{i}}\text{exp}\left(-\frac{E_{a,i}}{RT}\right)
    \label{eq:multiarrhenius}
\end{equation}
In the Multi-Arrhenius equation, the index $i$ refers to the $i$th set of Arrhenius parameters, which are consistent with those outlined in equation \ref{eq:arrhenius}.

\item[]\textbf{[3] Pdep-Arrhenius}-type expression

\vspace{0.5mm} 

Type 3, the Pdep-Arrhenius-type expression,  defines the rate coefficient $k(T, P)$ as temperature- and pressure-dependent, formulated through either Arrhenius (i.e., Type 1) or Multi-Arrhenius (i.e., Type 2) equations over multiple pressures (see Equation \ref{eq:pdeparrhenius}).
\begin{equation}
    \begin{split}
    k(T,P)& = A(P)\left({\frac{T}{T_{0}}}\right)^{n(P)}\text{exp}\left(-\frac{E_{a}(P)}{RT}\right)\\
    &\text{or}\\
    k(T,P)& = \sum_{i}A_i(P)\left({\frac{T}{T_{0,i}}}\right)^{n_i(P)}\text{exp}\left(-\frac{E_{a,i}(P)}{RT}\right)
    \label{eq:pdeparrhenius}
    \end{split}
\end{equation}
The rate coefficients are then determined by log scale interpolation between these expressions at each pressure. For example, the rate at an intermediate pressure $ P_1 < P < P_2 $ is computed as 
\begin{equation}
    \begin{split}
        \text{log } k(T, P) = &\text{log }k(T, P_1) \\
        &+\text{log}\left(\frac{k(T, P_2)}{k(T, P_1)}\right) \frac{\text{log}P-\text{log}P_1}{\text{log}P_2-\text{log}P_1}
    \label{eq:pdeparrhenius_interpolation}
    \end{split}
\end{equation}
In the Pdep-Arrhenius equation, $P$ refers to pressure [$Pa$]. $A(P)$ represents the pressure-dependent pre-exponential factor, $n(P)$ the pressure-dependent temperature exponent, and $E_0(P)$ the pressure-dependent activation energy in joules per mole [$J\cdot mol^{-1}$]. The unit of $A(P)$ depends on the reaction order—[$s^{-1}$] for first-order, [$m^3 \cdot mol^{-1} \cdot s^{-1}$] for second-order, and [$m^6 \cdot mol^{-2} \cdot s^{-1}$] for third-order reactions. The index $i$ refers to the $i$th set of Arrhenius parameters. For pressures beyond the specified range, that is, $P \leq P_{\text{lowest}}$ or $P \geq P_{\text{highest}}$, the rate-coefficient is determined by the Arrhenius equation at $P_{\text{lowest}}$ for lower, and at $P_{\text{highest}}$ for higher pressure values.
 
\item[]\textbf{[4] Multi-Pdep-Arrhenius}-type expression

\vspace{0.5mm} 

Type 4 is the Multi-Pdep-Arrhenius-type expression which defines the rate coefficient $k(T,P)$ as temperature- and pressure-dependent, formulated through both Arrhenius (i.e., Type 1) and Multi-Arrhenius (i.e., Type 2) equations (see Equation \ref{eq:multipdeparrhenius}).
\begin{equation}
    \begin{split}
    k(T,P)& = A(P)\left({\frac{T}{T_{0}}}\right)^{n(P)}\text{exp}\left(-\frac{E_{a}(P)}{RT}\right)\\
    &\text{and}\\
    k(T,P)& = \sum_{i}A_i(P)\left({\frac{T}{T_{0,i}}}\right)^{n_i(P)}\text{exp}\left(-\frac{E_{a,i}(P)}{RT}\right)
    \label{eq:multipdeparrhenius}
    \end{split}
\end{equation}

The rate coefficients are then determined by log scale interpolation between these expressions at each pressure as described in Equation~(\ref{eq:pdeparrhenius_interpolation}). In the Multi-Pdep-Arrhenius equation, all terms are consistent with those defined in the Pdep-Arrhenius expression (Type 3). Unlike the Pdep-Arrhenius-type expression, which formulates the rate coefficient exclusively as either Type 1 or Type 2, the Multi-Pdep-Arrhenius-type expression offers a less consistent approach. In this equation, the rate coefficient expression can vary among Type 1 and Type 2, thus presenting a mixed format.

\item[]\textbf{[5] Third-Body}-type expression

\vspace{0.5mm} 

Type 5 is the Third-Body-type expression whose kinetics simply introduces an inert third body to the rate expression as shown in Equation~(\ref{eq:thirdbody}).
\begin{equation}
    \begin{split}
    k(T,P) = &k_0(T)[M]\\
    &k_0(T) = A\left({\frac{T}{T_0}}\right)^n\text{exp}\left(-\frac{E_a}{RT}\right)\\
    &[M] = \frac{P}{RT}
    \label{eq:thirdbody}
    \end{split}
\end{equation}
In the Third-Body equation, $k_0(T)$ refers to the low-pressure limit temperature-dependent rate coefficient. Its unit depends on the reaction order—[$m^3 \cdot mol^{-1} \cdot s^{-1}$] for first-order, and [$m^6 \cdot mol^{-2} \cdot s^{-1}$] for second-order reactions. $[M]$ is the concentration of the bath gas [$mol \cdot m^{-3}$].

\item[]\textbf{[6] Lindemann}-type expression

\vspace{0.5mm} 

Type 6 is the Lindemann-type expression which qualitatively models the falloff behavior of pressure-dependent reactions as shown in Equation~(\ref{eq:Lindemann}).
\begin{equation}
    \begin{split}
    k(T,P) = &k_\infty(T)\left[\frac{P_r}{1+P_r}\right]\\[0.2cm]
    &P_r=\frac{k_0(T)}{k_\infty(T)}[M]\\[0.2cm]
    &k_0(T) = A_0\left({\frac{T}{T_{0,0}}}\right)^{n_0}\text{exp}\left(-\frac{E_{a,0}}{RT}\right)\\[0.2cm]
    &k_\infty(T) = A_{\infty}\left({\frac{T}{T_{0,\infty}}}\right)^{n_\infty}\text{exp}\left(-\frac{E_{a,\infty}}{RT}\right)\\
    \label{eq:Lindemann}
    \end{split}
\end{equation}
In the Lindemann equation, the Arrhenius expressions (i.e., Type 1) $k_0$ and $k_\infty$ represent the low-pressure and high-pressure limit kinetics, respectively. The units of $k_0$ and $k_\infty$ vary with reaction order: for first-order reactions, they are [$m^3 \cdot mol^{-1} \cdot s^{-1}$] and [$s^{-1}$], and for second-order reactions, [$m^6 \cdot mol^{-2} \cdot s^{-1}$] and [$m^3 \cdot mol^{-1} \cdot s^{-1}$], respectively.

\item[]\textbf{[7] Troe}-type expression

\vspace{0.5mm} 

Type 7 is the Troe-type expression which quantitatively models the falloff behavior of pressure-dependent reactions by introducing a broadening factor $F$ to the Lindemann equation, as shown in Equation \ref{eq:Troe}.
\begin{equation}
    \begin{split}
    k(T,P) = &k_\infty(T)\left[\frac{P_r}{1+P_r}\right]F\\[0.2cm]
    &P_r=\frac{k_0(T)}{k_\infty(T)}[M]\\[0.2cm]
    &k_0(T) = A_0\left({\frac{T}{T_{0,0}}}\right)^{n_0}\text{exp}\left(-\frac{E_{a,0}}{RT}\right)\\[0.2cm]
    &k_\infty(T) = A_{\infty}\left({\frac{T}{T_{0,\infty}}}\right)^{n_\infty}\text{exp}\left(-\frac{E_{a,\infty}}{RT}\right)\\
    \label{eq:Troe}
    \end{split}
\end{equation}
The broadening factor $F$ is computed via following Equation \ref{eq:Troe_F}
\begin{equation}
    \begin{split}
    \text{log}F = &\biggl\{1+\left[\frac{\text{log}P_r+c}{n-d(\text{log}P_r)+c}\right]^2\biggr\}^{-1}\text{log}F_{cent}\\[0.15cm] 
    &c = {-}0.4-0.67\cdot\text{log}F_{cent}\\[0.15cm] 
    &n = 0.75-1.27\cdot\text{log}F_{cent}\\[0.15cm] 
    &d = 0.14\\
    &F_{cent}= (1-\alpha)\text{exp}\left(\frac{-T}{T_3}\right)+\alpha\text{exp}\left(\frac{-T}{T_1}\right) \\ & \: \quad \quad + \text{exp}\left(\frac{-T_2}{T}\right)
    \label{eq:Troe_F}
    \end{split}
\end{equation}

In the Troe equation, the Arrhenius expressions (i.e., Type 1) $k_0$ and $k_\infty$ represent the low-pressure and high-pressure limit kinetics, respectively. The units of $k_0$ and $k_\infty$ vary with reaction order: for first-order reactions, they are [$m^3 \cdot mol^{-1} \cdot s^{-1}$] and [$s^{-1}$], and for second-order reactions, [$m^6 \cdot mol^{-2} \cdot s^{-1}$] and [$m^3 \cdot mol^{-1} \cdot s^{-1}$], respectively. Four parameters (i.e., $\alpha$, $T_1$, $T_2$, and $T_3$) are provided to calculate the broadening factor $F$.

\item[]\textbf{[8] Chebyshev}-type expression

\vspace{0.5mm} 

Type 8 is the Chebyshev-type expression which adopts the Chebyshev polynomial formulation as a means of fitting a wide range of complex $k(T,P)$ behavior as shown in Equation \ref{eq:Chebyshev}.
\begin{equation}
    \begin{split}
    \text{log}k(T,P) &= \sum_{t=1}^{N_T}\sum_{p=1}^{N_P}\alpha_{tp}\phi_t(\widetilde{T})\phi_p(\widetilde{P})\\[0.3cm] 
    &\widetilde{T} = \frac{2T^{-1}-T_{min}^{-1}-T_{max}^{-1}}{T^{-1}_{max}-T_{min}^{-1}}\\[0.3cm] 
    &\widetilde{P} = \frac{2\text{log}P-\text{log}P_{min}-\text{log}P_{max}}{\text{log}P_{max} - \text{log}P_{min}}\\[0.3cm]  
        &\boldsymbol{\alpha} =
        \begin{bmatrix}
            \alpha_{11} & \alpha_{12} & \cdots & \alpha_{1N_{P}}\\
            \alpha_{21} & \alpha_{22} & \cdots & \alpha_{2N_{P}}\\
            \vdots & \vdots & \ddots & \vdots \\
            \alpha_{N_{T}1} & \alpha_{N_{T}2} & \cdots & \alpha_{N_{T}N_{P}}\\
        \end{bmatrix}
    \label{eq:Chebyshev}
    \end{split}
\end{equation}
In the Chebyshev equation, $\alpha_{tp}$ are the constants defining the rate coefficient, and $\phi_n(x)$ is the Chebyshev polynomial of the first kind of degree $n$ evaluated at $x$. The first few Chebyshev polynomials of the first kind are described in Equation \ref{eq:Chebyshev_polynomials}.
\begin{equation}
    \begin{split}    
\phi_n(x)\,\rightarrow\:&\phi_1(x)=1\\
    &\phi_2(x)=x\\
    &\phi_3(x)=4x^3-3x\\
    &\phi_4(x)=8x^4-8x^2+1\\
    &\phi_5(x)=16x^5-20x^3+5x\\
    &\phi_6(x)=32x^6-48x^4+18x^2-1\\
    &\quad \qquad\vdots
    \end{split}
    \label{eq:Chebyshev_polynomials}
\end{equation}
$\widetilde{T}$ and $\widetilde{P}$ represent the reduced temperature and reduced pressures, respectively, mapping the ranges ($T_{min}, T_{max}$) and ($P_{min}, P_{max}$) to the interval (-1, 1). The Chebyshev rate expression is defined by the coefficient matrix $\boldsymbol{\alpha}$, comprising $\alpha_{tp}$, specific temperature and pressure ranges, typically involving 6 values for temperature (i.e., $N_T$=6) and 4 for pressure (i.e., $N_P$=4). It is important to note that Chebyshev polynomials are only defined within the interval (-1,1). Therefore, extrapolating rates beyond the defined temperature and pressure ranges is strongly discouraged, as the polynomials do not provide valid results outside these limits.
\end{itemize}

\subsection{1-D photochemical kinetic-transport atmospheric modeling using EPACRIS}\label{sec:epacris}

\begin{figure*}
    \centering    
    \includegraphics[width=1\textwidth]{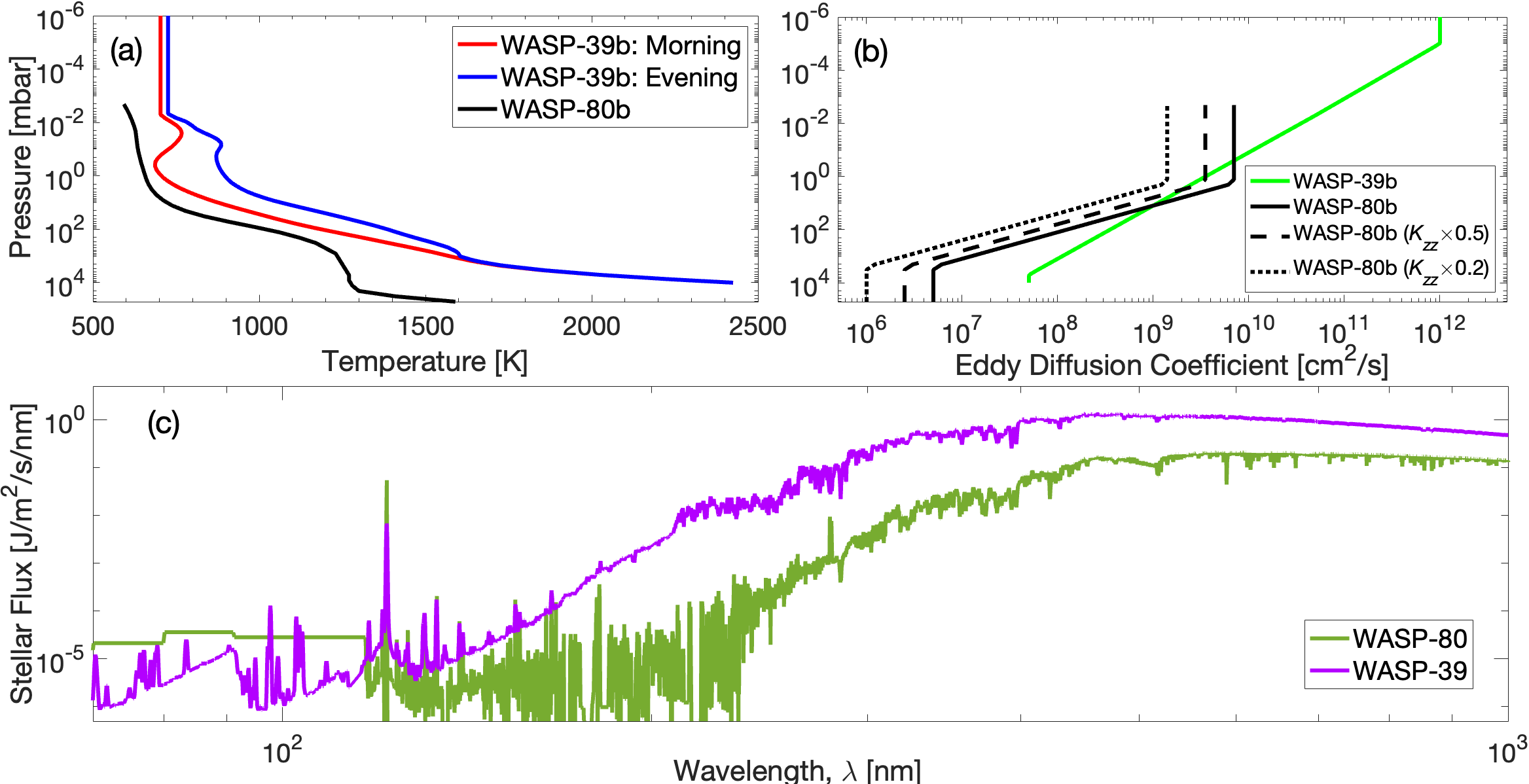}
     \caption{\footnotesize (a) Temperature-pressure profiles for the morning (red) and evening limbs (blue) for WASP-39~b, and for WASP-80~b (black) (b) eddy diffusion coefficient ($K_{zz}$) profile for WASP-39~b (lime), WASP-80~b (black solid line), 2$\times$ slower eddy diffusion coefficient profile for WASP-80~b (black dashed line), and 5$\times$ slower eddy diffusion coefficient profile for WASP-80~b (black dotted line), and (c) the stellar flux at the 1 AU distance for WASP-39 (purple) and WASP-80 (green). These input parameters are adopted from \cite{Tsai_2023, Bell-2023} except for the stellar spectra of WASP-80. The stellar flux of HD 85512 (K6V) at the 1 AU distance, adopted from the MUSCLES survey III \cite{Loyd_2016}, was used for WASP-80~b.}  
    \label{fig:epacris_inputs}
\end{figure*}

\begin{figure*}
    \centering
    \includegraphics[width=1.0\textwidth]{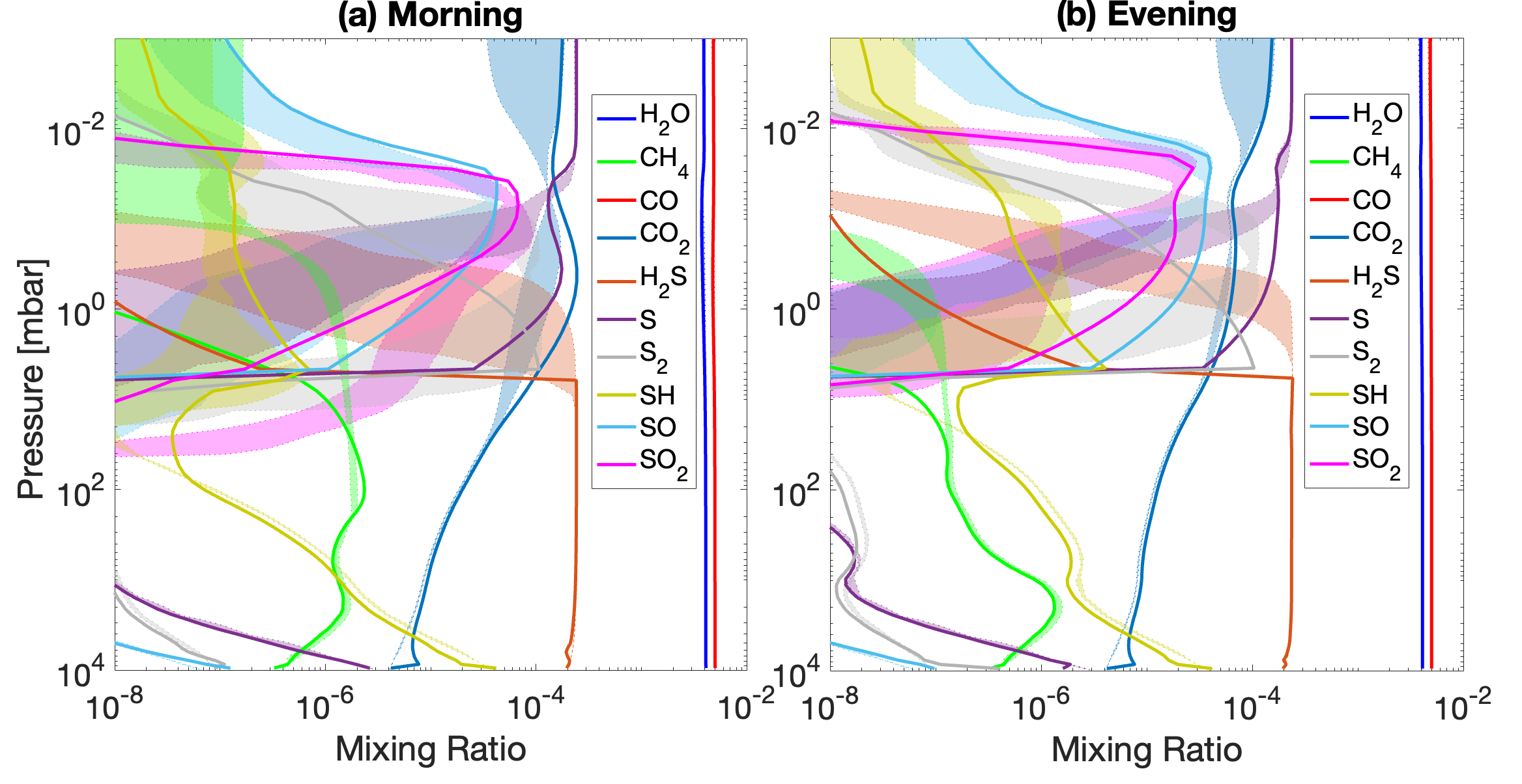}
    \caption{\footnotesize Comparison between the previously reported vertical molecular mixing ratio profile of major species simulated for the (a) morning and (b) evening terminators of WASP-39~b in \cite{Tsai_2023} (color-shaded areas) and those simulated from the current work using EPACRIS (solid lines). Each color indicates the corresponding species (\ce{SO2}: magenta, \ce{H2O}: blue, \ce{CH4}: green, \ce{CO}: red, \ce{CO2}: dark blue, \ce{H2S}: brown, \ce{S}: purple, \ce{S2}: grey, \ce{SH}: yellow, and \ce{SO}: light blue), and the color-shaded areas indicate the span enclosed by the photochemical models presented in \cite{Tsai_2023}.}
    \label{fig:wasp-39b_comparison}
\end{figure*}

After generating the chemical network by RMG for the conditions relevant to the \ce{H2}-dominated atmospheres of warm and hot Jupiters whose equilibrium temperature is 800--1500 K (Section \ref{sec:rmg}) and adapting it for EPACRIS using the RMG-EPACRIS adapter (Sections \ref{sec:sorting} and \ref{sec:RMG-EPACRIS_conversion}), we performed 1D photochemical kinetic-transport atmospheric modeling with EPACRIS to simulate the steady-state mixing ratio of chemical species in the atmospheres of WASP-39~b and WASP-80~b. The photochemical kinetic-transport module of EPACRIS was employed to calculate the steady-state chemical composition of WASP-39~b's atmosphere of each morning and evening terminator \citep[following][]{Tsai_2023} and that of WASP-80~b's atmosphere \citep[following][]{Bell-2023}, considering thermochemical equilibrium, vertical transport, and photochemical processes. We assumed cloud-free conditions and zero-flux boundary conditions. The temperature-pressure profiles (Figure~\ref{fig:epacris_inputs}a), the eddy diffusion coefficient profiles (Figure~\ref{fig:epacris_inputs}b), and the stellar spectra (Figure~\ref{fig:epacris_inputs}~c) are adopted from \cite{Tsai_2023, Bell-2023}. In the case of WASP-80~b, we used the stellar flux of HD 85512 (K6V) at the 1 AU distance, adopted from the MUSCLES survey III \citep{Loyd_2016}. It should be noted that the stellar spectrum significantly influences photolysis rates. From this, one can intuitively infer that the atmospheric chemistry of WASP-39~b is more significantly impacted by photochemistry compared to that of WASP-80~b based on Figure \ref{fig:epacris_inputs}c, which shows that the stellar flux on WASP-39b is 10 to 100 times stronger than that on WASP-80~b. We assumed atmospheric abundances of 10$\times$solar metallicity \citep{Lodders-2020} for both WASP-39~b \citep{Rustamkulov-2023, Tsai_2023} and WASP-80~b \citep{Bell-2023}. These choices facilitate a direct comparison between the EPACRIS-simulated WASP-39~b and WASP-80~b atmospheres and published results. After the model has converged and reached the steady state, we computed the synthetic transmission spectra of WASP-39~b and WASP-80~b based on the molecular mixing ratio profiles (Figure \ref{fig:wasp-39b_comparison} and \ref{fig:wasp-80b_comparison}), using the transmission spectra generation module of EPACRIS \citep{Hu_2013}, and compared the resulting transmission spectra with JWST observations \citep{Rustamkulov-2023, Powell_2024, Bell-2023}.

\section{Results and Discussions} \label{sec:Results_and_Discussions}
\subsection{WASP-39~b}
\subsubsection{Overall behavior of main sulfur-bearing species in the atmosphere of WASP-39b} \label{sec: sulfur_behaviors}

Figure \ref{fig:wasp-39b_comparison} shows the comparison between the previously reported vertical molecular mixing ratio profile of major species and those simulated from the current work using EPACRIS. As shown in Figure \ref{fig:wasp-39b_comparison}, the vertical mixing ratios of all species at pressures higher than $\sim10^2$ mbar are consistent across all five models, including EPACRIS and four others in \cite{Tsai_2023}. This consistency suggests that deep atmospheric chemistry of WASP-39b is primarily governed by thermal chemistry, which excludes photochemistry, and aligns with thermochemical equilibrium as depicted in Figure A\ref{fig:comparison_WASP-39b_equil}. The minor variations in the behaviors of SH and \ce{S2} are primarily attributed to differences in the Gibbs free energy of these species. For example, VULCAN \citep{Tsai_2017, Tsai_2021}, one of the models used in \cite{Tsai_2023}, utilizes thermodynamic parameters for \ce{SH} and \ce{S2} from \cite{Burcat-newnasa-9} and \cite{Mcbride-2002-nasa}. These sources calculated NASA polynomials based on the data from \cite{Mcbride-2002-nasa}. In contrast, RMG uses thermodynamic parameters for \ce{SH} whose NASA polynomials are calculated by \cite{Song-2017}, which are based on the data from \cite{Shiell-2000}, and for \ce{S2} from the NIST-JANAF table \citep{chase-thermo}. It is evident that each parameter in photochemical modeling inherently contains a certain amount of uncertainties. Among these, in general, rate coefficients and thermodynamic parameters are the most significant contributors to the uncertainties of chemical kinetic models. Therefore, assessing the model's sensitivity to thermodynamic parameters as well as rate coefficients is of paramount importance when it comes to the preciseness of atmospheric photochemical modeling.

At pressure less than $\sim10^2$ mbar (i.e., where photochemistry and vertical mixing become more significant), the EPACRIS simulations (solid lines in Figure \ref{fig:wasp-39b_comparison}) of the main sulfur species (\ce{H2S}, \ce{SH}, \ce{S2}, \ce{S}, \ce{SO}, \ce{SO2}) align closely with the models in \cite{Tsai_2023}. The peak mixing ratios of these species are generally within an order of magnitude of each other for both morning and evening terminators, as illustrated in Figures \ref{fig:wasp-39b_comparison}a and b. Although these models represent the atmospheric steady state, making it challenging to track the time-dependent chemical evolution, some insights can still be gleaned. For instance, in the deep atmosphere (around $\sim10^4$ mbar), thermochemically favored \ce{H2S} (hydrogen sulfide) is the dominant sulfur-bearing species up to a pressure of about 10 mbar (at temperatures above 900 K), for both terminators. In addition to \ce{H2S}, sulfur monohydride (SH) is the second most abundant sulfur species in the deep atmosphere. Between 5--8 mbar, \ce{H2S} rapidly transitions to \ce{S2} and \ce{S}, with \ce{SO} and \ce{SO2} also present. Above 1 mbar, atomic sulfur (\ce{S}) becomes the dominant sulfur species, with \ce{SO} and \ce{SO2} peaking the maximum mixing ratio at $\sim10^{-1}-10^{-2}$ mbar for both the morning ([SO]$_{max}$ = $\sim$ 42 ppm, [\ce{SO2}]$_{max}$ = $\sim$68 ppm) and evening ([SO]$_{max}$ = $\sim$40 ppm, [\ce{SO2}]$_{max}$ = $\sim$28 ppm) terminators.

One notable difference between the current model (i.e., EPACRIS) and the models previously reported in \cite{Tsai_2023} lies in the amount of \ce{SO2} levels predicted between 0.5 to 10 mbar for both the morning and evening terminators of WASP-39b (see Figure \ref{fig:wasp-39b_comparison}a and b). In the morning terminator, \cite{Tsai_2023} predicts \ce{SO2} mixing ratios above 1 ppm (i.e., $10^{-6}$) starting from around 10 mbar. In contrast, EPACRIS indicates this level is reached at around 2 mbar, a higher altitude (Figure \ref{fig:wasp-39b_comparison}a). Conversely, for the evening terminator, \cite{Tsai_2023} suggests \ce{SO2} exceeds 1 ppm starting from 0.5 mbar, whereas EPACRIS shows this occurring at around 3 mbar, a lower altitude (Figure \ref{fig:wasp-39b_comparison}b). 
This increased formation of \ce{SO2} on the evening side compared to the morning side is explained in Section~\ref{sec: OH_source}. 
It has to be noted that the mixing ratio prediction at around 1 mbar is potentially significant for JWST observations in transmission, especially considering the NIRSpec mode's primary probing range of 0.1 to 2 mbar \citep{Rustamkulov-2023}. 

Another notable difference is the amount of \ce{CH4} levels predicted in the atmosphere above $P\sim$10 mbar for both the morning and evening terminators (see Figure \ref{fig:wasp-39b_comparison}a and b). According to the thermochemical equilibrium vertical mixing ratio profile of \ce{CH4} as shown in Figure A\ref{fig:comparison_WASP-39b_equil}), \ce{CH4} mixing ratio at around 10 mbar should be more than 100 ppm in the morning terminator, and more than 1 ppm in the evening terminator, but this is not the case since the quenching kinetics starts to happen around 100 mbar for both the morning and evening terminator, followed by photodissociation of \ce{CH4} at higher altitude ($P\sim$1 mbar), as shown in Figure \ref{fig:wasp-39b_comparison}. However, on top of this, the current model shows a more depleted \ce{CH4} level compared to previous modeling studies in \cite{Tsai_2023}, which indicates additional scavenging reactions for \ce{CH4} sources (e.g., \ce{CH3}). According to the rate analysis, the reactions
\begin{equation}
    \begin{split}
        \ce{CH3}+\ce{S}&\rightarrow\ce{CH3S}\\
        \ce{CH3}+\ce{SH}&\rightarrow\ce{CH3SH}
    \end{split}
    \label{eqn: sulfur_impacting_methane_sinks}
\end{equation}
lead to continuous reactions forming various carbon and sulfur-bearing species, ultimately resulting in the depletion of methane. 
As pointed out in \cite{Tsai_2023}, sulfur can affect other nonsulfur species including \ce{CH4} (see Extended Data Fig. 6 in \cite{Tsai_2023}), and the additional sulfur species listed in Table \ref{tbl: newspecies} (e.g., \ce{CH2SH} is not included in VULCAN or KINETICS, both of which were used in \cite{Tsai_2023}) and related reactions (e.g., $\ce{CH3}+\ce{S}\rightarrow\ce{CH3S}$ is not included in VULCAN, \cite{Tsai_2023}) in RMG-generated chemical networks are attributed to this additional \ce{CH4} depletion in the upper atmospheres of WASP-39~b. This rapid drop of \ce{CH4} abundance at around 1 mbar matches with the rapid appearance of \ce{S} and \ce{S2} originated from the dissociation of \ce{H2S} (see Figure \ref{fig:wasp-39b_comparison}). 

\begin{figure}
    \centering    
    \includegraphics[width=0.5\textwidth]{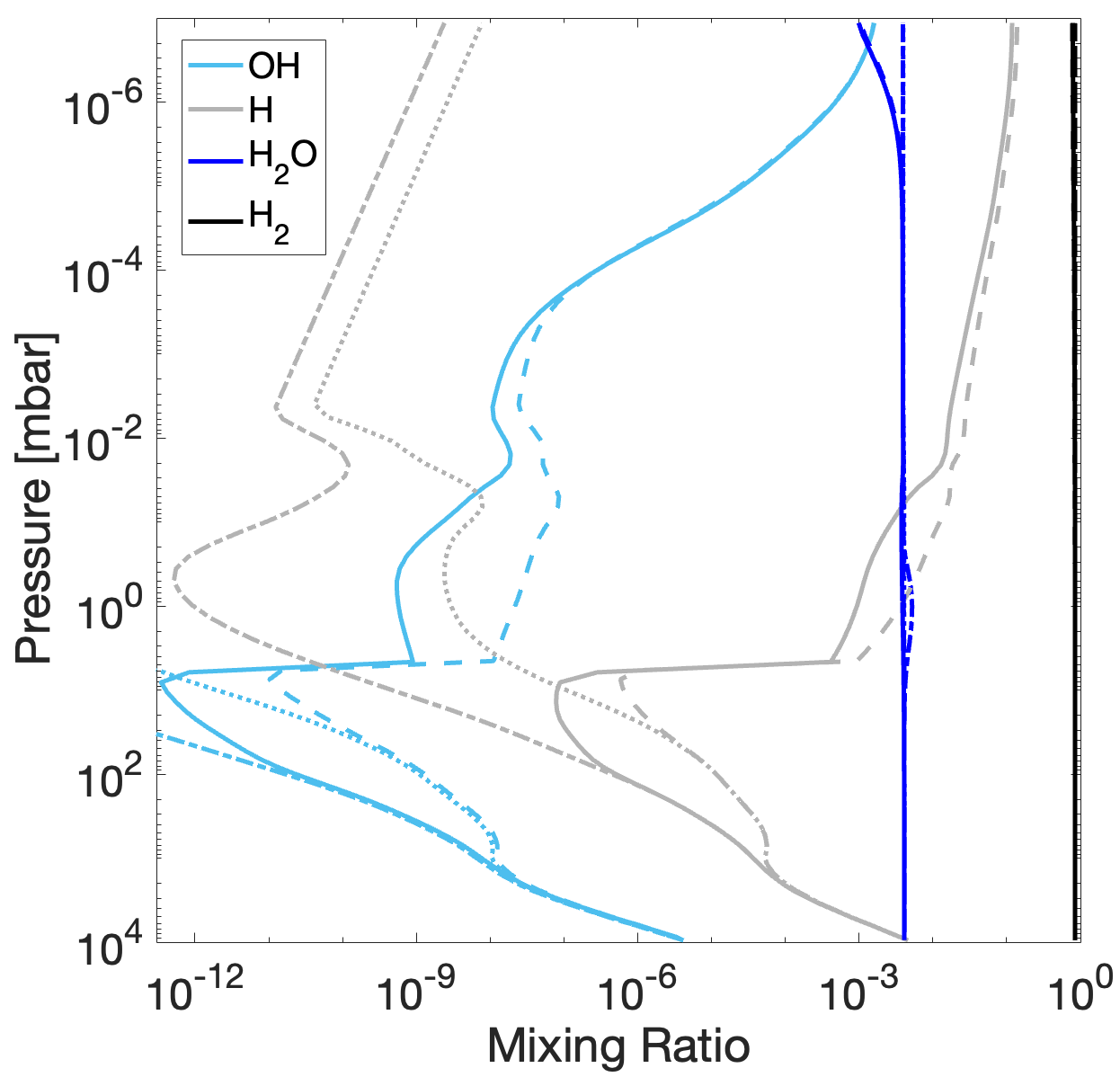}
     \caption{\footnotesize The vertical molecular mixing ratio profile of several species (\ce{H2}, \ce{H2O}, \ce{H}, \ce{OH}) simulated for the morning (solid lines) and evening (dashed lines) terminators of the WASP-39~b atmosphere. The thermochemical equilibrium vertical molecular mixing ratios are also indicated by the dash-dotted lines (morning terminator) and dotted lines (evening terminator). Although not shown in the figure, the thermochemical equilibrium vertical molecular mixing ratios for the morning and evening terminators of OH show similar patterns to those of H radicals with the $\sim$6 orders of magnitude smaller amplitude. Each color indicates the corresponding species (\ce{OH}: light blue, \ce{H}: grey, \ce{H2O}: blue, and \ce{H2}: black).}  
    \label{fig:H_OH_ratio}
\end{figure}

\subsubsection{Various origins of H and OH radicals for the formation of \ce{SO2} in the atmosphere of WASP-39~b} \label{sec: OH_source}
Generally, sulfur dioxide (\ce{SO2}) at high altitudes ($P\sim10^{-2}-0.5$ mbar) is more prevalent at the cooler morning terminator of WASP-39b, whereas at lower altitudes ($P\sim0.5-10$ mbar), \ce{SO2} is more abundant at the hotter evening terminator. To understand this distribution pattern, it is crucial to track the origin of oxidizers (i.e., OH and H radicals), since \ce{SO2} in the atmosphere of WASP-39~b is mainly produced by successive oxidation of sulfur species originating from the deep atmosphere hydrogen sulfide (\ce{H2S}) as pointed out in the previous modeling work of \cite{Tsai_2023}.

\begin{figure*}
    \centering    
    \includegraphics[width=\textwidth]{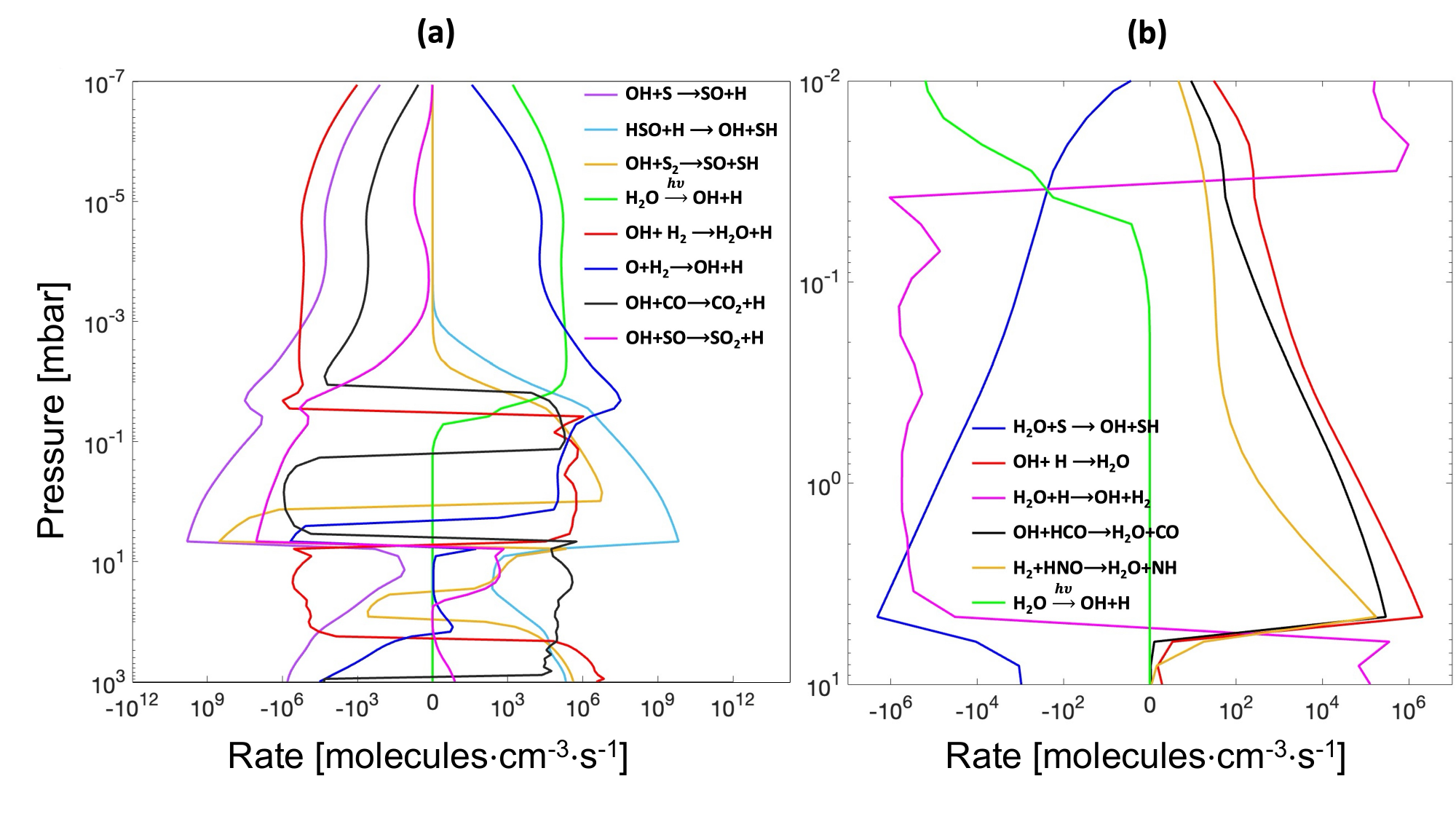}
     \caption{\footnotesize The rate-pressure profile of dominant reactions involving specific chemical species in the morning terminator: \textbf{(a)} rates of reactions involving OH radicals presented at pressure ranges between $10^{-7}$--$10^3$ mbar. Among 318 OH-involved reactions, only 8 dominant reactions are shown here for readability. Each color of the solid lines indicates a corresponding reaction. Negative values in the rate indicate that the reaction acts as a sink for OH species, while positive values indicate the reaction serves as a source for OH species. The rate-pressure profile for the evening terminator exhibited behavior similar to that of the morning terminator, with the primary difference being in the amplitude of the rates (the evening rates are in general slightly faster within a factor of 2--3). Due to this similarity, the evening profile is not separately illustrated; \textbf{(b)} reactions involving \ce{H2O}, presented at pressure ranges between $10^{-2}$--10 mbar. Among 124 \ce{H2O}-involved reactions, only 6 dominant reactions are shown here for readability. Each solid line color corresponds to a specific reaction, with the colors representing the six rates in descending order from largest to smallest. Negative values in the rate indicate the reaction consumes \ce{H2O}, while positive values indicate the reaction that produces \ce{H2O}. As mentioned in the main text, in the upper atmosphere, \ce{H2O} photolysis (lime) serves as a major source for OH species, while \ce{H2O} + S$\rightarrow$OH + SH reaction (blue) in the middle atmosphere at a pressure of 1--10 mbar serves as a major source for OH.
}  
    \label{fig:OH_H2O_rates}
\end{figure*}

As illustrated in Figure \ref{fig:H_OH_ratio}, the vertical molecular mixing ratios of both OH and H radicals display similar patterns, largely due to the dissociation of \ce{H2O} into H and OH. This also indicates that \ce{H2O} is a main source for OH as well as H. However, their magnitudes differ, with additional sources of H such as \ce{H2} and \ce{H2S} contributing to these variations. The morning and evening vertical mixing ratio profiles of these species (i.e., OH and H radicals), as shown by the solid (morning) and dashed (evening) lines in Figure \ref{fig:H_OH_ratio}, cannot be fully explained by thermochemical equilibrium (dash-dotted line for the morning and dotted line for the evening) alone. This discrepancy indicates that a combination of thermal chemistry, photochemistry, and vertical mixing influences these behaviors. Further analysis, as shown in Figure~\ref{fig:OH_H2O_rates}a, indicates that the origins of OH radicals change with altitude (or pressure), suggesting a complex interplay of atmospheric processes at different levels. As shown in Figure \ref{fig:OH_H2O_rates}a, from the upper atmosphere at around $10^{-7}$ mbar down to $10^{-3}$ mbar, the reactions
\begin{equation}
    \ce{H2O}\xrightarrow{\text{h$\nu$}}\ce{OH}+\ce{H}
\end{equation}
\begin{equation}    \ce{OH}+\ce{H2}\rightarrow\ce{H2O}+\ce{H}
\end{equation}
serve as the major OH source and sink, respectively. (rates $\sim\pm2\times10^5$ [molecules/cm$^3$/s])

Then in between $10^{-3}-10^{-1}$ mbar, the reactions
\begin{equation}
    \ce{O}+\ce{H2}\rightarrow\ce{OH}+\ce{H}
\end{equation}
\begin{equation}
    \ce{OH}+\ce{S}\rightarrow\ce{SO}+\ce{H}
\end{equation}
serve as the major OH source and sink, respectively. (rates $\sim\pm3\times10^7$ [molecules/cm$^3$/s]). And as shown in Figure \ref{fig:OH_H2O_rates}a, at the pressure between $10^{-1}-10^{1}$ mbar, the reactions
\begin{equation}
    \ce{HSO}+\ce{H}\rightarrow\ce{OH}+\ce{SH}
\end{equation}
\begin{equation}
    \ce{OH}+\ce{S}\rightarrow\ce{SO}+\ce{H}
\end{equation}
serve as the major OH source and sink, respectively (rates $\sim\pm6\times10^9$ [molecules/cm$^3$/s]). The reaction rates get larger with decreasing altitude since molecular number density [molecules/cm$^3$] gets larger with decreasing altitude (i.e., increasing pressure $\propto$ number density). This interconversion of OH and H radicals is rapid, leading to the formation of a combined H + OH chemical group whose relative ratio remains constant under specific atmospheric conditions, as depicted in Figure~\ref{fig:H_OH_ratio}. 

As highlighted previously, water vapor (\ce{H2O}) is a primary source of these radicals. Rate analysis involving \ce{H2O} (Figure~\ref{fig:OH_H2O_rates}~b) shows that the formation of OH and H, crucial for \ce{SO2} production, primarily occurs through two distinct reactions in different atmospheric regimes. In the upper atmosphere, at pressures below $10^{-2}$ mbar, \ce{H2O} photolysis is the predominant reaction (see Figure~\ref{fig:OH_H2O_rates}a and b. Conversely, in the middle atmosphere, within the pressure range of 1 to 10 mbar, the interaction between \ce{H2O} and sulfur radicals (originated from \ce{H2S}) becomes increasingly significant (as shown in Figure~\ref{fig:OH_H2O_rates}b). These interactions can be summarized by the following reactions: 
\begin{equation}
    \begin{split}
        \ce{H2S}&\rightarrow\ce{H2}+\ce{S}\\
        \ce{H2O}+\ce{S}&\rightarrow\ce{OH}+\ce{SH}\\
        \ce{SH}+\ce{S}&\rightarrow\ce{H}+\ce{S2}\\
        \ce{S2}&\xrightarrow{\text{h$\nu$}}\ce{S}+\ce{S}\\
        \hline
        \text{Net}: \ce{H2S}+\ce{H2O}&\rightarrow\ce{H2}+\ce{OH}+\ce{H}+\ce{S}\\
    \end{split}
    \label{eqn: H2S_conversion_into_S}
\end{equation}

This summarized scheme is consistent with the rapid increase in \ce{H}, \ce{OH} around 10 mbar, as shown in Figure~\ref{fig:H_OH_ratio}, and the rapid depletion of \ce{H2S} along with a rapid increase in \ce{S} at the same pressure level, as shown in Figure \ref{fig:wasp-39b_comparison}a and b. It is also noteworthy that the thermochemical equilibrium mixing ratio of \ce{H2O} at pressures between 1--10 mbar is slightly higher than its vertical mixing ratio of \ce{H2O} (see Figure~\ref{fig:H_OH_ratio}). The transport rate (vertical mixing) in this region (i.e., $P\sim$1--10 mbar) was at least 2 orders of magnitude slower than the total loss rates. This suggests that vertical transport is not the predominant factor for the straight feature of \ce{H2O} vertical mixing ratio profile at this region. Instead, this straight feature indicates that the decreased amount of \ce{H2O} in this region is being converted into H and OH, corroborating the above reaction scheme.

If we combine the $T-P$ profiles (see Figure \ref{fig:epacris_inputs}a) with OH vertical mixing ratio of the morning and evening terminator (solid and dashed lines, respectively, in Figure \ref{fig:H_OH_ratio}), we can see the positive correlation between the OH (and H) radical mixing ratios and temperature within the $10^{-1}-10^1$ mbar pressure range. Notably, in this range (i.e., $10^{-1}-10^1$ mbar), the morning terminator consistently exhibits temperatures $\sim$200 K lower than the evening terminator. As described earlier, the origin of the lower atmospheric OH is more thermally driven chemistry while less \ce{H2O} photolysis-driven, and thus sensitive to temperatures. As a result, within the $10^{-1}-10^1$ mbar pressure range, OH radicals are more than an order of magnitude more abundant in the hotter evening terminator compared to the cooler morning terminator, as illustrated in Figure \ref{fig:H_OH_ratio}. This increased OH abundance leads to more \ce{SO2} formation at lower altitudes ($P\sim0.5-10$ mbar) through OH-aided successive oxidation.

\subsubsection{Theoretical transmission spectra of the atmosphere of WASP-39~b generated by EPACRIS}\label{subsec:vs_observations_wasp-39b}
\begin{figure*}
    \centering    
    \includegraphics[width=1\textwidth]{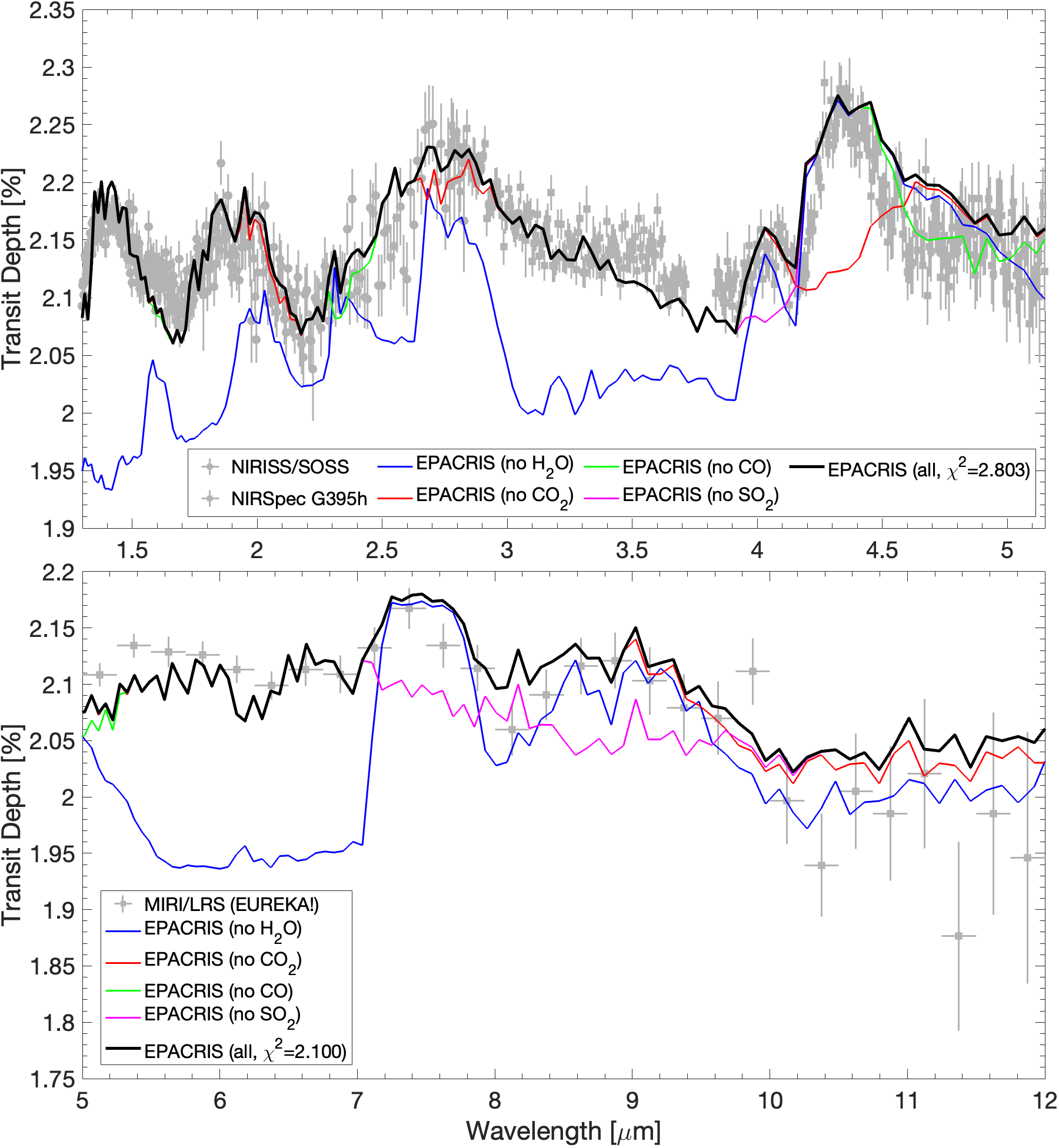}
     \caption{\footnotesize Comparison between the terminator-averaged theoretical transmission spectra generated by EPACRIS (solid lines) and the JWST observations: (top) NIRISS/SOSS and NIRSpec/G395H data \citep{Feinstein-2023, Alderson-2023} (grey circle symbol points with error-bars indicate NIRISS/SOSS data, while grey square symbol points with uncertainties indicate NIRSpec G395h) and (bottom) MIRI/LRS data \citep{Powell_2024} (grey square symbol points with uncertainties). The uncertainties are 1$\sigma$ standard deviations. The reduced $\chi^2$ value for the near-infrared region was calculated against the NIRISS/SOSS and NIRSpec/G395H data, while the reduced $\chi^2$ value for the mid-infrared region was calculated against the MIRI/LRS data. Each color represents a spectrum generated by excluding specific species: blue for no \ce{H2O}, red for no \ce{CO2}, green for no CO, magenta for no \ce{SO2}, and black for all species included.}  
    \label{fig:transmission_spectra_wasp-39b}
\end{figure*}

Figure \ref{fig:transmission_spectra_wasp-39b} shows the comparison between the averaged theoretical transmission spectra generated by EPACRIS and JWST observations of WASP-39~b \citep{Alderson-2023, Feinstein-2023, Powell_2024}.
The EPACRIS-predicted transmission spectra are broadly consistent with the 
near- and mid-infrared observations. The model accurately captures the features of \ce{H2O} (blue), \ce{CO2} (red), and \ce{SO2} (magenta). While the \ce{CO} (green) feature at 4.5--5 $\mu m$ appears to be overpredicted, considering the uncertainties, the overall agreement between the model predictions and observations is still considered decent. 
The predicted transmission spectra generated by the previous photochemical networks, assuming 10$\times$ solar metallicity as discussed in \cite{Tsai_2023}, align well with the NIRSpec/G395H spectra \citep{Feinstein-2023}. However, they overpredicted the transit depth in the 7--8 $\mu m$ wavelength range, which corresponds to the dominant \ce{SO2} absorption feature when compared to the MIRI data reported by \cite{Powell_2024}. Consequently, using the same photochemical networks resulted in the lowest $\chi^2$ value of 2.51 when assuming 7.5$\times$ solar metallicity, while the $\chi^2$ for 10$\times$ solar metallicity was 2.91 \citep{Powell_2024}. In contrast, the predicted transmission spectra using the photochemical network from the current study showed an even lower $\chi^2$ of 2.10. This network matches well with both NIRISS/SOSS, NIRSpec/G395H, and MIRI/LRS spectra without the need to vary solar metallicity. 
As highlighted in the previous Section \ref{sec: sulfur_behaviors}, the vertical mixing ratios of species in the 0.1--2 mbar pressure range are probed by JWST observations in transmission. Differences in the photochemical network can lead to significantly varied results in the vertical mixing ratio profiles, thus influencing the predicted transmission spectra. 

\subsection{WASP-80~b}
\subsubsection{Upper atmospheric chemistry affected by the deep interior thermochemistry and quenching kinetics} \label{sec: wasp-80b_chemistry}

\begin{figure*}
    \centering
    \includegraphics[width=1\textwidth]{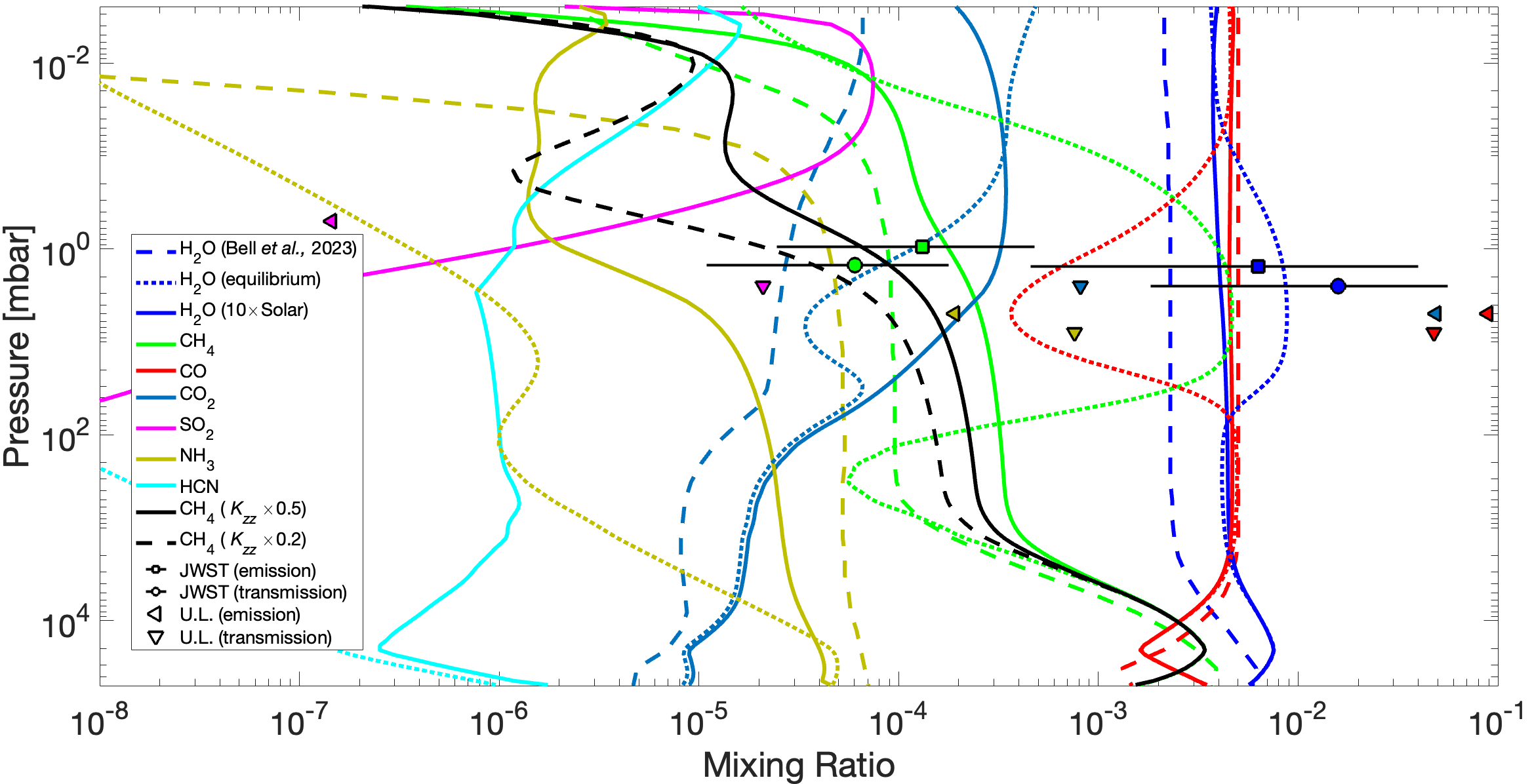}
    \caption{\footnotesize Comparison of the vertical molecular mixing ratio profiles for WASP-80~b's atmosphere: previous JWST observations (\cite{Bell-2023}, symbols and dashed lines) versus current EPACRIS simulations (solid lines for 1D photochemical kinetic-transport modeling; dotted lines for thermochemical equilibrium). Square symbols with uncertainties indicate the emission data, and circle symbols with uncertainties indicate the transmission spectra. Left-pointing triangle symbols indicate the upper limit (U.L.) values determined from emission spectra, while down-pointing triangle symbols indicate the upper limit values determined from transmission spectra. Each color indicates the corresponding species (\ce{SO2}: magenta, \ce{H2O}: blue, \ce{CH4}: green, \ce{CO}: red, \ce{CO2}: dark blue, \ce{NH3}: yellow, and \ce{HCN}: cyan). The dark solid and dashed lines are the predicted \ce{CH4} volume mixing ratio when using 2$\times$ slower $K_{zz}$ profile and 5$\times$ slower $K_{zz}$ profile in Figure \ref{fig:epacris_inputs}b, respectively. Although not shown, the \ce{H2O} volume mixing ratio using 2$\times$ slower $K_{zz}$ profile is almost identical to the \ce{H2O} volume mixing ratio using the original $K_{zz}$ profile of WASP-80~b.}
    \label{fig:wasp-80b_comparison}
\end{figure*}

WASP-80~b's equilibrium temperature ($T_{eq} = 825 K$, \cite{triaud-2015}) is approximately 300 K cooler than WASP-39~b's ($T_{eq}$ = 1116 K, \cite{faedi-2011}). This suggests that transport-induced quenching, where the lifetime of chemical species becomes longer relative to the vertical mixing timescale, could play a more significant role in the cooler atmosphere of WASP-80~b compared to the hotter WASP-39~b. Figure \ref{fig:wasp-80b_comparison} illustrates that all major species, including \ce{H2O}, \ce{CO}, \ce{CH4}, \ce{NH3}, and \ce{HCN}, originate from the deep interior (the quenching point is at around $P=10^3$ mbar) and are transported to the upper atmosphere, where some species such as \ce{CH4} and \ce{NH3} undergo photodissociation. Notably, the model-predicted \ce{H2O} volume mixing ratio aligns well with the JWST observations \citep{Bell-2023} as well as \ce{CH4} prediction. Although other major species such as \ce{NH3}, \ce{HCN}, \ce{CO2}, \ce{CO}, and \ce{SO2} were not constrained from the observations, all these model-predicted mixing ratios fall within the upper limit values derived from emission and transmission spectra except an \ce{SO2} upper limit determined from emission spectra (see magenta left-pointing triangle symbol in Figure \ref{fig:wasp-80b_comparison}).

As expected, the \ce{CH4} mixing ratio as well as other species (e.g., \ce{NH3} and \ce{HCN}) formed from deep interior thermochemistry are sensitive to quenching kinetics. As depicted in Figure \ref{fig:wasp-80b_comparison}, the black solid and dashed lines represent the \ce{CH4} volume mixing ratio using an eddy diffusion coefficient ($K_{zz}$) profiles that are 2 times slower and 5 times slower than the $K_{zz}$ profile adopted from \cite{Bell-2023}, respectively (see Figure \ref{fig:epacris_inputs}b). When using these slower $K_{zz}$ profiles, the predicted \ce{CH4} mixing ratio becomes more consistent with the observational constraints while not changing the predicted \ce{H2O} mixing ratio, and \ce{NH3} and \ce{HCN} mixing ratios shift to the lower mixing ratios, indicating a shift of the deep interior quenching point toward lower pressure (see Figure \ref{fig:wasp-80b_comparison}). Consequently, more detailed constraints on other species such as \ce{CO2}, \ce{NH3}, \ce{CO}, and \ce{SO2} are required to precisely describe the WASP-80~b's atmospheric chemistry. Despite uncertainties in metallicity and the $K_{zz}$ profile, the current model aligns reasonably well with the observational data, providing valuable insights into the atmospheric behavior of this exoplanet.

\subsubsection{Detailed chemistry and newly suggested deep-interior nitrogen incorporation pathway} \label{sec: newly_suggested_deep_interior_chemistry}

As shown in Figure \ref{fig:deep-chemistry}a, the deep-interior \ce{CO}--\ce{CH4} conversion is 
\begin{equation}
    \begin{split}
        \ce{CO}+\ce{H}&\xrightarrow{\text{M}}\ce{HCO}\\
        \ce{HCO}+\ce{H2}&\rightarrow\ce{CH2O}+\ce{H}\\
        \ce{CH2O}+\ce{H}&\xrightarrow{\text{M}}\ce{CH2OH}\\
        \ce{CH2OH}+\ce{H2}&\rightarrow\ce{CH3OH}+\ce{H}\\
        \ce{CH3OH}&\xrightarrow{\text{M}}\ce{CH3}+\ce{OH}\\
        \ce{CH3}+\ce{H2}&\rightarrow\ce{CH4}+\ce{H}\\
        \hline
        \text{Net}: \ce{CO}+3\ce{H2}&\rightarrow\ce{CH4}+\ce{H2O},\\
    \end{split}
    \label{eqn: CO_CH4_conversion}
\end{equation}
with M representing any third-body molecule. This scheme is identical to the scheme (1) in \cite{Moses_2016}. 
The \ce{SO2} formation mechanism in the upper atmosphere predicted in the model was similar to the \ce{SO2} formation in the atmosphere of WASP-39~b previously described in Section \ref{sec: OH_source}. The \ce{CO2} formation mechanism was the combination of deep interior \ce{CO2} formation and the additional oxidation of \ce{CO} through \ce{OH} radicals at the atmosphere above $P\sim$ 1 mbar.

Figure \ref{fig:deep-chemistry}b visualizes the \ce{N2} to \ce{NH3} to \ce{HCN} conversion pathway in the deep interior. This pathway mostly resembles those detailed in \cite{Moses_2016}, with a key difference in the initial incorporation of nitrogen from \ce{N2} into species like \ce{NH3} and \ce{HCN}. The well-known \ce{N2}$\rightarrow$\ce{NH3} route, as outlined in \cite{Moses_2016}, involves multiple hydrogenation steps starting with \ce{N2} activation by a hydrogen atom to form \ce{NNH}, ultimately yielding \ce{NH3} as following:
\begin{equation}
    \begin{split}
        \ce{N2}+\ce{H}&\xrightarrow{\text{M}}\ce{NNH}\\
        \ce{NNH}+\ce{H2}&\rightarrow\ce{N2H2}+\ce{H}\\
        \ce{N2H2}+\ce{H}&\rightarrow\ce{NH}+\ce{NH2}\\
        \ce{NH}+\ce{H2}&\rightarrow\ce{NH2}+\ce{H}\\
        2 (\ce{NH2}+\ce{H2}&\rightarrow\ce{NH3}+\ce{H})\\
        \ce{2H}&\xrightarrow{\text{M}}\ce{H2}\\
        \hline
        \text{Net}: \ce{N2}+\ce{3H2}&\rightarrow\ce{2NH3}\\
    \end{split}
    \label{eqn: Moses_2016_N2_NH3}
\end{equation}

Despite this scheme being included in the RMG-generated chemical network for hot Jupiter exoplanet atmospheres described in Section \ref{sec:rmg}, RMG suggests a different dominant deep-interior pathway for \ce{N2}$\rightarrow$\ce{NH3} conversion that initiates with \ce{N2} directly interacting with two \ce{H2} molecules to form \ce{N2H2}, eventually leading to \ce{NH3} (highlighted in Figure \ref{fig:deep-chemistry}b) as following scheme:
\begin{equation}
    \begin{split}
        \ce{N2}+\ce{H2}+\ce{H2}&\rightarrow\ce{N2H2}+\ce{H2}\\
        \ce{N2H2}+\ce{H2}&\rightarrow\ce{N2H4}\\
        \ce{N2H4}&\rightarrow\ce{2NH2}\\
        \ce{2NH2}+\ce{2H2}&\rightarrow\ce{2NH3}+\ce{2H}\\
        \ce{2H}&\xrightarrow{\text{M}}\ce{H2}\\
        \hline
        \text{Net}: \ce{N2}+\ce{3H2}&\rightarrow\ce{2NH3}.\\
    \end{split}
    \label{eqn: rmg_N2_NH3}
\end{equation}

\begin{figure}
    \centering
    \includegraphics[width=0.45\textwidth]{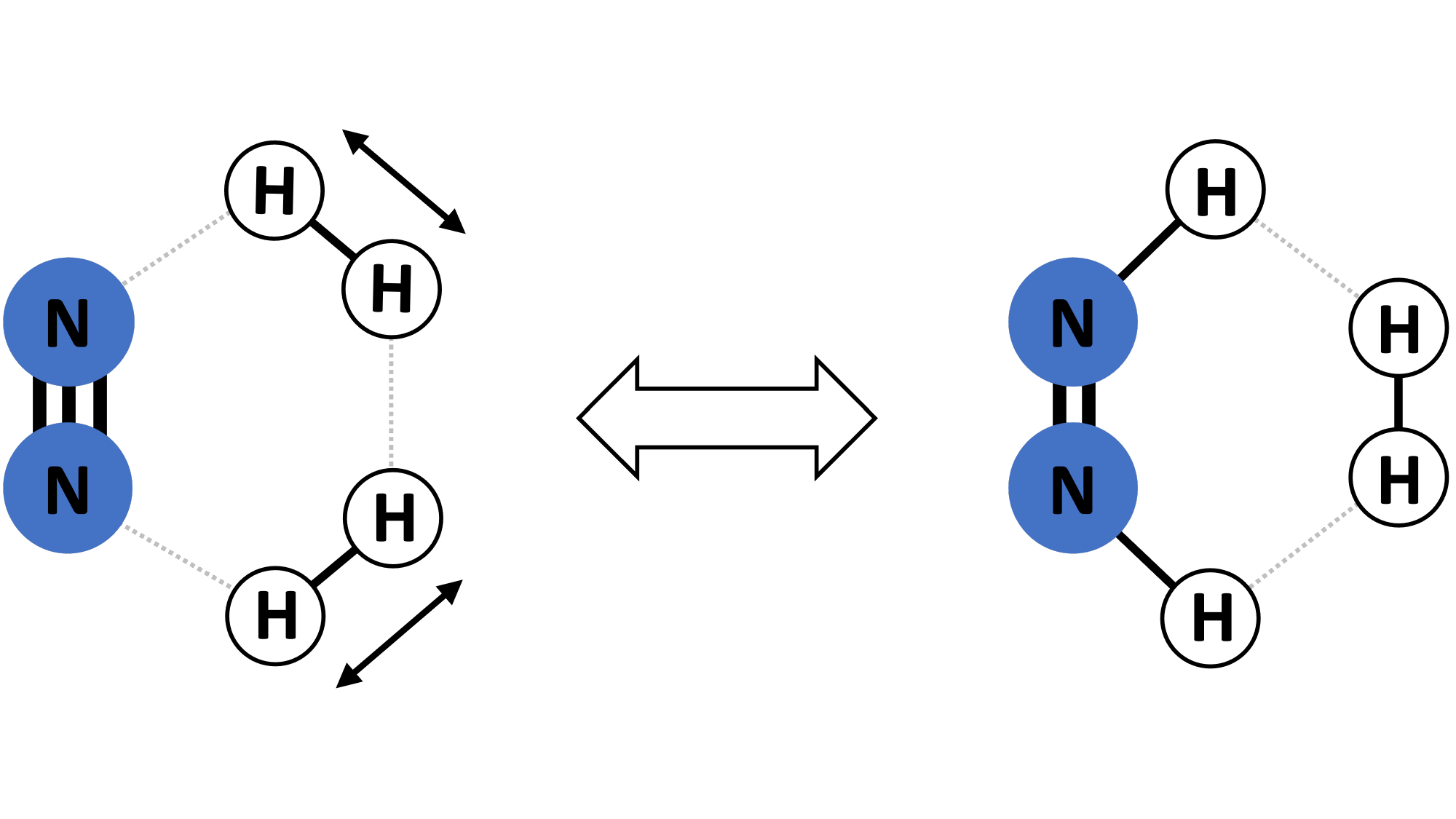}
    \caption{\footnotesize A schematic diagram that visualizes the first reaction: \ce{N2}+\ce{H2}+\ce{H2}$\rightarrow$\ce{N2H2}+\ce{H2} in Equation~\ref{eqn: rmg_N2_NH3}. From the left side, \ce{N2} reacts with two \ce{H2} molecules to form \textit{cis-}\ce{N2H2} and \ce{H2} on the right side.}
    \label{fig:dihydrogen-activated_Nitrogen_fixation}
\end{figure}

\begin{figure}
    \centering
    \includegraphics[width=0.47\textwidth]{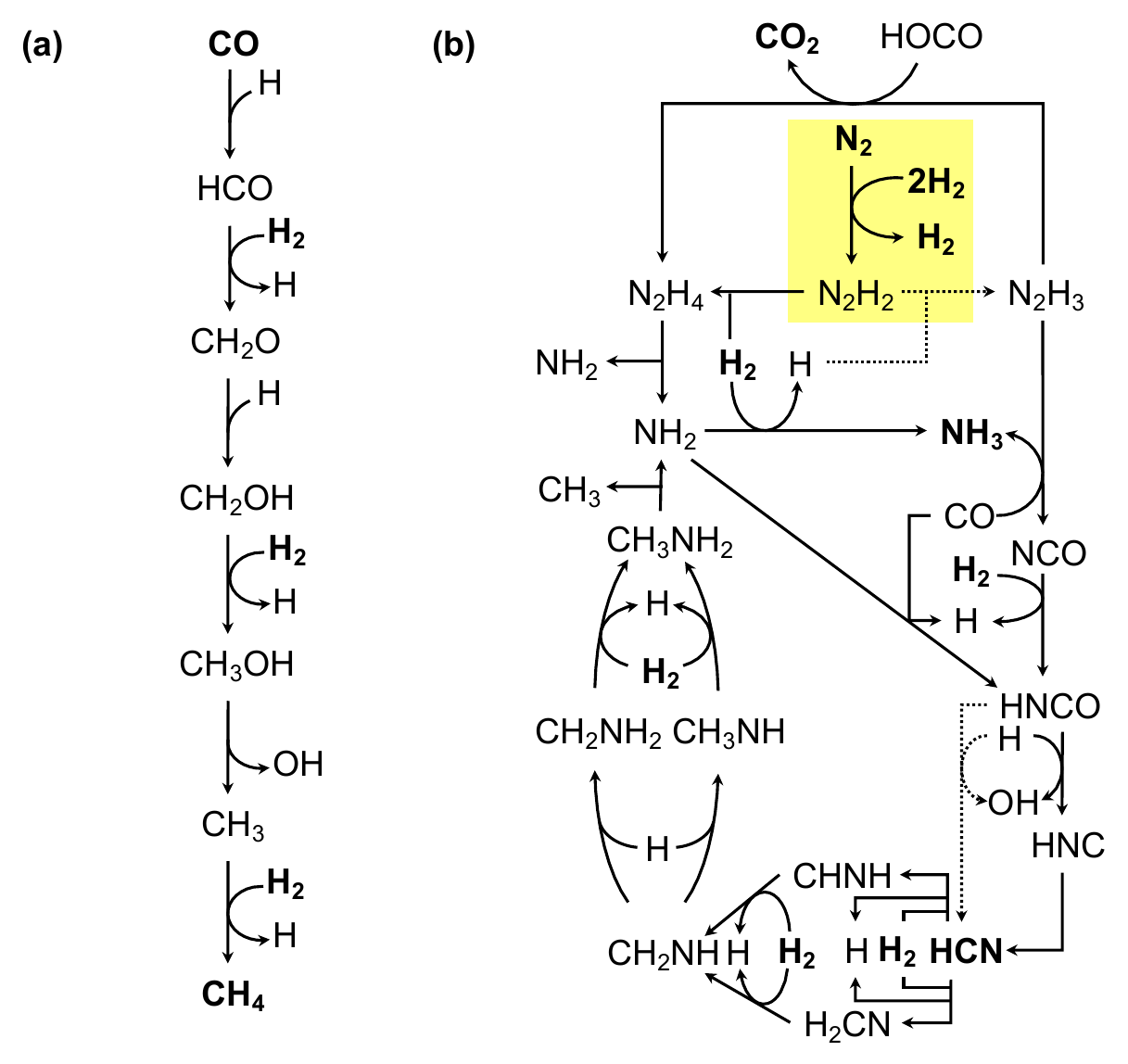}
    \caption{\footnotesize Model-predicted major reaction pathways at deep-interior ($P\sim3\times10^4$mbar) for (a) \ce{CO}--\ce{CH4} conversion, and (b) \ce{N2}--\ce{NH3}--\ce{HCN} conversion. Bold characters represent major species with a mixing ratio above 1 ppm. Dashed pathways indicate directions at least three times slower than other branching reactions (e.g., \ce{N2H2} branching into \ce{N2H3} is 3$\times$ slower than into \ce{N2H4}). The yellow highlighted region in panel (b) indicates a newly suggested initial nitrogen incorporation path in the RMG-generated chemical network.}
    \label{fig:deep-chemistry}
\end{figure}

In this scheme, \ce{N2} is activated by two \ce{H2} molecules, forming \textit{cis}-\ce{N2H2} (see Figure~\ref{fig:dihydrogen-activated_Nitrogen_fixation}). The transition state for this reaction was first identified by \cite{Asatryan-2010} using the CBS-QB3 level of theory. In contrast to the simple bimolecular addition of \ce{N2}+\ce{H2} with an initial barrier of 125 kcal/mol (calculated at G2M(MP2)//MP2/6-31G** level of theory by \cite{hwang-2003}), this dihydrogen-activated nitrogen fixation has a relatively much lower barrier of about 77 kcal/mol \citep{Asatryan-2010}. While the initial barrier in Scheme \ref{eqn: Moses_2016_N2_NH3} is lower (17.1 kcal/mol, \cite{hwang-2003}), subsequent steps, such as \ce{NNH}+\ce{H2}$\rightarrow$\ce{N2H2}+\ce{H}, face much higher barriers (42 kcal/mol, \cite{hwang-2003}), and Scheme \ref{eqn: Moses_2016_N2_NH3} has to go through one more additional elementary reaction compared to Scheme \ref{eqn: rmg_N2_NH3}. We tested the sensitivity of the photochemical model to the newly suggested reaction by intentionally disabling it. We found that under the conditions of WASP-80~b's atmosphere, the overall mixing ratio did not show significant changes. 
However, under certain favorable deep-atmospheric conditions (e.g., hot and high-pressure conditions), this newly suggested \ce{N2} incorporation step could lead to significantly more formation of \ce{NH3} or \ce{HCN} species.

In this study, the rate coefficient for the \ce{N2}+\ce{2H2}$\rightarrow$\ce{N2H2}+\ce{H2} is considered as a high-pressure limit, calculated via conventional transition state theory. However, as this termolecular reaction involves three actual reactants (unlike the usual third body [$M$] considered as an unreacted agent), it is inherently pressure-dependent and entropically less favorable. Despite these, such reactions are likely viable under the high pressures characteristic of the deep-interior chemistry in hot Jupiter atmospheres, underscoring the importance of its inclusion for accurate nitrogen incorporation modeling. This case highlights the substantial advantages of systematic, computer-aided automatic chemical network generation, which can reveal previously overlooked chemical pathways and provide detailed insights into exoplanet atmospheric chemistry. 

As shown in Figure \ref{fig:deep-chemistry}b, the \ce{NH3}$\rightarrow$HCN conversion scheme at pressures between $10^3-10^4$ mbar is 
\begin{equation}
    \begin{split}
        \ce{NH3}+\ce{H}&\rightarrow\ce{NH2}+\ce{H2}\\
        \ce{CH4}+\ce{H}&\xrightarrow{\text{M}}\ce{CH3}+\ce{H2}\\
        \ce{NH2}+\ce{CH3}&\xrightarrow{\text{M}}\ce{CH3NH2}\\
        \ce{CH3NH2}+\ce{H}&\rightarrow\ce{CH3NH}\text{ (or } \ce{CH2NH2})+\ce{H2}\\
        \ce{CH3NH} \text{ (or } \ce{CH2NH2})&\xrightarrow{\text{M}}\ce{CH2NH}+\ce{H}\\
        \ce{CH2NH}+\ce{H}&\rightarrow\ce{H2CN} \text{ (or } \ce{CHNH})+\ce{H2}\\
        \ce{H2CN} \text{ (or } \ce{CHNH})+\ce{H}&\rightarrow\ce{HCN}+\ce{H2}\\
        \ce{H2}&\xrightarrow{\text{M}}\ce{H}+\ce{H}\\
        \hline
        \text{Net}: \ce{NH3}+\ce{CH4}&\rightarrow\ce{HCN}+\ce{3H2}.\\
    \end{split}
    \label{eqn: rmg_NH3_HCN}
\end{equation}

This pathway mostly resembles those detailed in \cite{Moses_2016}, with additional species (i.e., \ce{CH3NH} and \ce{CHNH} are newly included species by RMG) and reactions (i.e., \ce{CH3NH2}+\ce{H}$\rightarrow$\ce{CH3NH}+\ce{H2}, \ce{CH3NH}$\xrightarrow{\text{M}}$\ce{CH2NH}+\ce{H}, \ce{CH2NH}+\ce{H}$\rightarrow$\ce{CHNH}+\ce{H2}, and \ce{CHNH}+\ce{H}$\rightarrow$\ce{HCN}+\ce{H2}).

\subsubsection{Theoretical transmission spectra of the atmosphere of WASP-80~b generated by EPACRIS}\label{subsec:vs_observations__wasp-80b}

\begin{figure*}
    \centering    
    \includegraphics[width=1\textwidth]{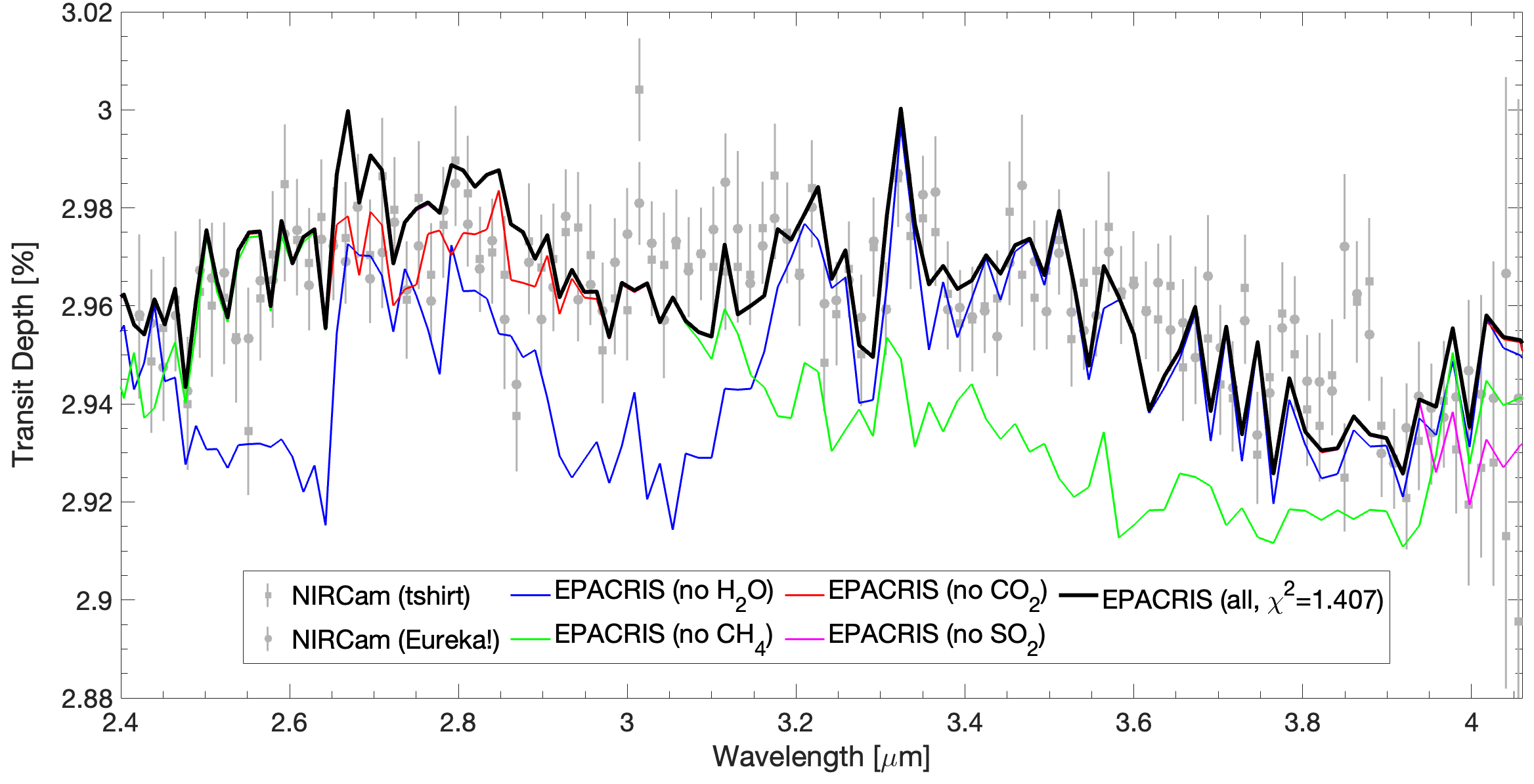}
     \caption{\footnotesize Comparison between the theoretical transmission spectra generated by EPACRIS (solid lines) and the JWST observations: NIRCam t-shirt (grey square symbol points with uncertainties) and Eureka! (grey circle symbol points with uncertainties) reductions by \cite{Bell-2023}. The uncertainties are 1$\sigma$ standard deviations. The reduced $\chi^2$ value of 1.407 was calculated against the NIRCam tshirt! reduction data. Although not shown, the reduced $\chi^2$ value against the NIRCam Eureka! reduction data was 1.748. Each color represents a spectrum generated by excluding specific species: blue for no \ce{H2O}, red for no \ce{CO2}, green for no \ce{CH4}, magenta for no \ce{SO2}, and black for all species included.}  
    \label{fig:transmission_spectra_wasp-80b}
\end{figure*}

Figure \ref{fig:transmission_spectra_wasp-80b} compares EPACRIS-generated theoretical transmission spectra with the JWST NIRCam observations of WASP-80~b \citep{Bell-2023}. The EPACRIS prediction aligns well with the NIRCam data ($\chi^2$=1.748 for Eureka! and 1.407 for tshirt), particularly in capturing \ce{H2O} (blue) and \ce{CH4} (green) features identified in \cite{Bell-2023}. Due to NIRCam's wavelength range limitations for \ce{CO2} and \ce{SO2} detection, comparing these species with model predictions is challenging. However, our model indicates that future JWST NIRSpec/G395H observations could potentially confirm the presence of \ce{CO2} and \ce{SO2}. Additionally, the spectral feature near 3 $\mu m$ could signify \ce{NH3} or \ce{HCN} presence, both anticipated to exceed 1 ppm at pressures below 1 mbar (Figure \ref{fig:wasp-80b_comparison}). This observation underscores the need for more detailed exploration in this wavelength range, potentially through repeated observations. Overall, similar to the WASP-39~b case, there is a decent agreement between the model predictions and the observational data on the WASP-80b atmosphere.

\subsection{Future applications and limitations of the current study}\label{subsec:future implication}
The current framework—automatic reaction mechanism generation coupled with one-dimensional (1-D) photochemical kinetic-transport modeling—has numerous potential applications for tools used in studying (exo)planetary atmospheres. One application is in the realm of climate modeling. Accounting for disequilibrium chemistry is essential in climate modeling. Consequently, reducing the size of the photochemical network becomes crucial, especially since two-dimensional (2-D) or three-dimensional (3-D) climate modeling is computationally intensive. This challenge amplifies in the context of general circulation modeling (GCM).

To address this, it is important to retain major chemical species that significantly impact climate structure while pruning less significant species from the chemical network to enhance computational efficiency. The current framework can offer substantial benefits to climate modeling and GCMs by eliminating unimportant species and reactions. Pruning can be achieved by adjusting features in the Reaction Mechanism Generator (RMG), such as increasing the user-specified error tolerance, $\epsilon$, or limiting the total number of atoms. However, this process involves a trade-off between minimizing the network size and maintaining the precision of the chemical network, necessitating a balance between these two aspects.

Moreover, several areas within the current framework could be improved. For instance, since RMG is primarily developed for simulating combustion chemistry, it lacks photochemical reactions in its library and does not account for vertical mixing processes (e.g., molecular diffusion and eddy diffusion). This omission might lead to significant gaps in identifying atmospherically important chemical species and reactions involving photons and vertical mixing. Therefore, incorporating photochemistry and physical processes in reaction mechanism generation is a critical area for future study.

\section{Conclusions} \label{sec:conclusions}

In this study, we have developed a new framework for exoplanet atmospheric photochemical modeling. This framework, for the first time, integrates a rate-based automatic chemical network generator (RMG) with a 1-D photochemical kinetic-transport atmospheric simulator, forming the chemistry module of EPACRIS. We first constructed the reaction network specifically tailored for the atmosphere of \ce{H2}-dominated hot Jupiter whose equilibrium temperature is 800--1500 K, and then incorporated this chemical network into EPACRIS for one-dimensional photochemical kinetic-transport modeling. 
Our model results generally align with previous studies of WASP-39~b by \cite{Tsai_2023}, particularly in capturing the photochemical production of OH radicals from \ce{H2O} photolysis in the upper atmosphere and the formation of \ce{SO2} through successive oxidation by OH and H radicals. A key difference between our study and previous models is the predicted \ce{SO2} abundance in the middle atmosphere (pressure range of 0.5–10 mbar). Our results indicate a higher \ce{SO2} formation at the warmer evening terminator compared to the cooler morning terminator. This discrepancy is attributed to an increased presence of sulfur species-oxidizers (OH and H radicals), predominantly generated from thermally-driven reactions between sulfur-bearing species (such as S or \ce{S2}) and \ce{H2O} within this middle atmosphere. The predicted transmission spectrum of the atmosphere of WASP-39~b based on our model was compared to the JWST NIRISS/SOSS, NIRSpec/G395H, and MIRI/LRS observations \citep{Feinstein-2023, Alderson-2023, Powell_2024}, showing good consistency and capturing \ce{H2O}, \ce{CO2}, and \ce{SO2} spectral features for a $10\times$Solar metallicity atmosphere. 
Our model result of the WASP-80~b shows that the deep interior chemistry and vertical mixing dominate the general atmospheric chemistry, with predicted concentrations of \ce{CH4}, \ce{H2O}, \ce{CO}, \ce{NH3}, \ce{HCN}, and \ce{SO2} exceeding 1 ppm at pressures below 1 mbar. Utilizing RMG, we identified a dominant, previously overlooked reaction for the initial nitrogen incorporation (\ce{N2}+2\ce{H2}$\rightarrow$\ce{N2H2}+\ce{H2}), significant in high-pressure environments like deep interior atmospheres. Such use of RMG can unveil new reactions within chemical networks, potentially leading to the discovery of novel species in (exo)planetary atmospheres. The predicted transmission spectrum of the atmosphere of WASP-80~b based on our model was compared to the JWST NIRCam observation reported by \cite{Bell-2023}, showing good consistency and capturing \ce{H2O} and \ce{CH4} spectral features.
This new approach not only provides the 1-D photochemical kinetic-transport modeling of (exo)planetary atmospheres with unprecedented efficiency and preciseness but also the applicability to diverse atmospheric conditions (e.g., from \ce{H2}-dominated to \ce{H2O}-dominated atmospheres), enabling us to more effectively and precisely predict and interpret the vast amounts of data from upcoming JWST observations of various exoplanet atmospheric conditions.

\section*{Acknowledgements}
The authors thank Shang-min Tsai, Julianne Moses, Sean Jordan, and Diana Powell for the discussions on the choice of reaction rate coefficients involving \ce{H2S} thermal dissociation. The authors thank Taylor Bell for providing the WASP-80~b observation data, and Diana Powell for providing the WASP-39~b observational data. This research work was carried out at the Jet Propulsion
Laboratory, California Institute of Technology, under a
contract with the National Aeronautics and Space Administration.
This research work was funded by the Caltech-JPL President's and Director's Research and Development Fund. © 2024. California Institute of Technology. Government sponsorship acknowledged.
%

\vspace{5mm}


\software{EPACRIS \citep{Hu_2012, Hu_2013, Hu_2014, Hu_2019}, RMG \citep{Gao_2016,rmg-v3, RMG-database, RMG-developers}}


\appendix
\setcounter{figure}{0} 

\section{Thermochemical equilibrium in the atmosphere of WASP-39b}\label{sec:appendix_a}
\begin{figure*}
    \renewcommand{\figurename}{Figure A}
    \centering    
    \includegraphics[width=\textwidth]{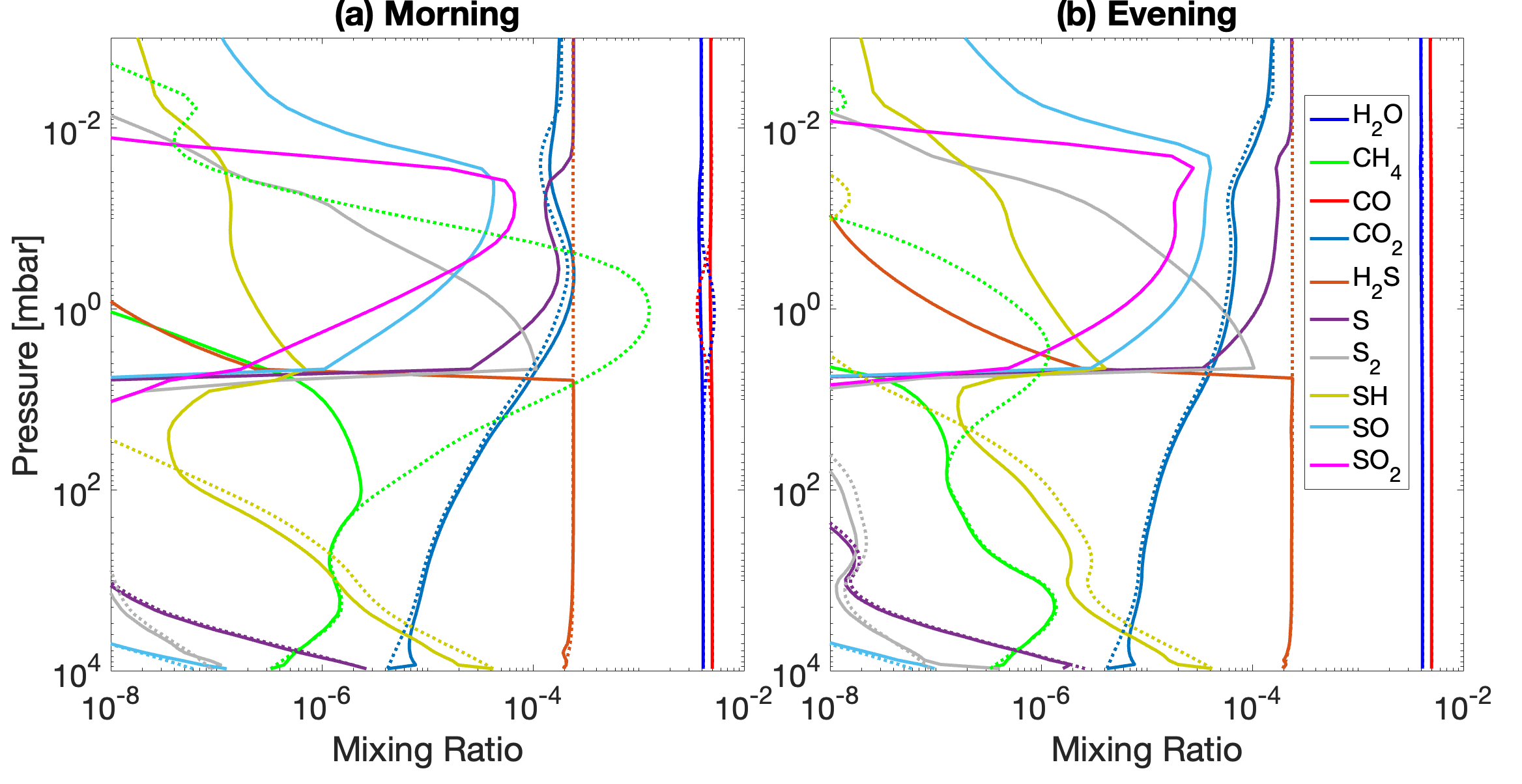}
     \caption{\footnotesize Comparison between the thermochemical equilibrium vertical molecular mixing ratio profile of major species simulated for the (a) morning and (b) evening terminators (dashed lines) and photochemical steady-state vertical molecular mixing ratio profile of major species simulated from the current work using EPACRIS (solid lines). Each color indicates the corresponding species (\ce{SO2}: magenta, \ce{H2O}: blue, \ce{CH4}: green, \ce{CO}: red, \ce{CO2}: dark blue, \ce{H2S}: brown, \ce{S}: purple, \ce{S2}: grey, \ce{SH}: yellow, and \ce{SO}: light blue)}  
    \label{fig:comparison_WASP-39b_equil}
\end{figure*}

Figure A\ref{fig:comparison_WASP-39b_equil} presents a comparison between the thermochemical equilibrium and photochemical steady-state vertical molecular mixing ratios of major species. These profiles were both simulated in the current study using EPACRIS.

\bibliography{reference}{}
\bibliographystyle{aasjournal}



\end{document}